\documentclass[11pt, letterpaper]{article}
\usepackage{cite} 
\usepackage[utf8]{inputenc}
\usepackage{amsmath,amssymb}
\usepackage{graphicx}
\usepackage{subfig}
\usepackage{here}
\usepackage{float}
\usepackage{multirow}
\usepackage[margin=0.75in]{geometry}
\usepackage{boldline}
\usepackage{xcolor}
\usepackage{authblk}
\usepackage{siunitx}
\usepackage{booktabs}
\usepackage{caption}
\usepackage{comment}
\usepackage[percent]{overpic}
\usepackage{mathtools}
\usepackage{hyperref}
\DeclareMathAlphabet{\mathpzc}{OT1}{pzc}{m}{it}
\newcommand{\vev}[1]{\langle\Omega|#1|\Omega\rangle}

\renewcommand{\Im}{\mathrm{Im}}

\newcommand{\glueFourD}{\big\langle \alpha G^{2} \big\rangle}
\newcommand{\glueSixD}{\big\langle g^{3} G^{3} \big\rangle}

\newcommand{\rom}[1]{\uppercase\expandafter{\romannumeral #1\relax}}

\newcommand{\eg}{\textit{e.g.,}}
\title{Bottom-Charm $[bc]$ Diquark Masses from QCD Laplace Sum-Rules}

\author[1]{T.\ de Oliveira\thanks{tdeoliveira@nycu.edu.tw}}
\author[7]{D.\ Harnett\thanks{
harnettd1@douglascollege.ca
}
}
\author[3]{J.\ Ho\thanks{Jason.Ho@dordt.edu}}
\author[4,2]{R.T.\ Kleiv\thanks{Robin.Kleiv@tru.ca}}
\author[5]{Siyuan\ Li\thanks{Siyuan.Li@uni-mainz.de}}
\author[6]{N.\ Loewen\thanks{Nathan.Loewen2@student.ufv.ca}}
\author[6]{A.\ Palameta\thanks{Alexander.Palameta@ufv.ca}}
\author[2]{T.G.\ Steele\thanks{Tom.Steele@usask.ca}}

\affil[1]{Institute of Physics, National Yang Ming Chiao Tung University, 30010 Hsinchu, Taiwan}
\affil[2]{Department of Physics and Engineering Physics\\ University of Saskatchewan\\ Saskatoon, SK, S7N 5E2, Canada}
\affil[3]{Department of Physics\\ Dordt University\\ Sioux Center, Iowa, 51250, USA}
\affil[4]{Department of Physical Sciences (Physics)\\ Thompson Rivers University\\ Kamloops, BC, V2C~0C8, Canada}
\affil[5]{Institut f\"ur Kernphysik,
 Johannes Gutenberg-Universit\"at  Mainz,  D-55128 Mainz, Germany}
\affil[6]{Department of Physics\\ University of the Fraser Valley\\ Abbotsford, BC, V2S 7M8, Canada}
\affil[7]{
Department of Physics, Douglas College, New Westminster, BC, V3L 5B2, Canada
}

\begin{document}
\maketitle
\begin{abstract}

\noindent Constituent mass predictions for bottom-charm $[bc]$ color-antitriplet diquarks are generated using QCD Laplace sum rules. We calculate the diquark correlator within the operator product expansion to next-to-leading-order (NLO), including terms proportional to the four- and six-dimensional gluon and six-dimensional quark condensates. Existing analytic expressions for the resulting multi-scale master integrals are insufficient for these calculations, so we utilize numerical $D$-dimensional loop integration techniques. 
The sum rule analyses stabilize in all four channels yielding constituent masses in GeV of: scalar ($0^+$) $(5.60\pm0.56)$, pseudoscalar ($0^-$) $(6.04\pm0.60)$, axial vector ($1^+$) $(5.60\pm0.56)$ and vector ($1^-$) $(5.99\pm0.60)$.
These constituent mass results are discussed in comparison with values used in diquark models of tetraquark hadronic states. We also summarize QCD sum rule predictions for diquark masses for all possible quark-flavor combinations, providing QCD-based criteria for assessing the diquark mass parameters used in constituent diquark models of hadrons.
\end{abstract}

\section{Introduction}\label{I}

Since the discovery of the $\chi_{c1}(3872)$~\cite{Belle:2003nnu,CDF:2003cab,D0:2004zmu,BaBar:2004iez,LHCb:2011zzp}, experiments have identified more than fifty candidates for exotic hadrons containing heavy ({\it i.e.}, charm and/or bottom) quarks~\cite{Lebed:2016hpi}. Examples include the closed-flavor $T_{cc\overline{cc}}(6900)$~\cite{LHCb:2020bwg,CMS:2023owd,ATLAS:2023bft}, as well as the open-flavor states $T^\ast_{\overline{cs}0}(2870)$~\cite{LHCb:2020pxc}, $T_{b\bar{s}}(5568)^+$~\cite{D0:2017qqm,D0:2016mwd}, and $T_{cc}(3875)^+$~\cite{LHCb:2021vvq,LHCb:2021auc}. The respective minimal valence contents of these states are $\bar{c}\bar{c}cc$, $\bar{c}\bar{s}ud$, $b\bar{s}ud$, and $cc\bar{u}\bar{d}$, hence they are all compelling tetraquark candidates. 

Given these discoveries, the $T_{bc}$ tetraquark systems represent a natural frontier for experimental and theoretical investigations.   
LHCb $T_{bc}$ discovery prospects have been examined in Ref.~\cite{Feng:2026rle}, and $T_{bc}$ properties have been studied using a variety of theoretical approaches. In particular, $T_{bc}$ tetraquarks have been studied via 
QCD sum rules ~\cite{Chen:2013aba,Agaev:2019kkz,Wang:2019xzt,Tang:2019nwv,Ozdem:2024dbq}, Bethe-Salpeter methods~\cite{Wang:2025apq}, quark models~\cite{Karliner:2017qjm,Carames:2018tpe,Wu:2024zbx},
and heavy-quark symmetry~\cite{Eichten:2017ffp}.
Lattice QCD studies of the $T_{bc}^{ud}$ are ongoing~\cite{Francis:2018jyb,Meinel:2022lzo,Radhakrishnan:2024ihu} (see also the review in Ref.~\cite{Francis:2024fwf} and references therein).

Two main pictures used to understand the internal substructure of tetraquarks are the molecular and compact scenarios\footnote{Models combining both scenarios have also been proposed (see {\it e.g.} Ref.~\cite{Esposito:2016itg}).}. The molecular scenario is motivated by the fact that many tetraquark masses are in very close proximity to two-meson 
thresholds, suggesting that those states could be weakly bound meson pairs. An alternative is the compact scenario, wherein tetraquarks are pictured as being composed of a pair of strongly interacting colored diquarks~\cite{Jaffe:2004ph,Anselmino:1992vg}. Treating the diquark and anti-diquark as constituents, various models can then be used to determine tetraquark properties. These include chiral quark models~\cite{Pepin:1996id,Tan:2020ldi}, Type I and Type II diquark models \cite{Maiani:2004vq,Maiani:2005pe,Maiani:2013nmn,Lebed:2016yvr,Maiani:2021tri,Maiani:2014aja,Lee:2009rt}, the chromomagnetic interaction model~\cite{Cheng:2020wxa}, dynamical quark models \cite{Vijande:2006jf,Giron:2021sla,Mutuk:2026bhq}, relativized diquark models \cite{Anwar:2018sol,Ferretti:2020ewe,Bedolla:2019zwg,Anwar:2017toa,Ferretti:2019zyh}, relativistic quark models \cite{Ebert:2005nc,Ebert:2007rn,Ebert:2008kb,Ebert:2010af,Lu:2020rog,Faustov:2021hjs}, 
 Regge relations and string models \cite{Song:2023izj},
and the diquark effective Hamiltonian model \cite{Shi:2021jyr}. The compact scenario has also been applied to pentaquarks, where diquarks are used in a variety of models (see e.g., \cite{Shi:2021wyt,Zhu:2015bba,Giron:2021fnl}). The constituent diquark mass is a crucial parameter in any compact hadron model utilizing diquarks. In many cases, compact models estimate the diquark mass using phenomenological methods.

The constituent diquark mass can be calculated using QCD sum rules, an approach pioneered in Ref.~\cite{Dosch1989} for $J^P=0^+$ light diquarks. Subsequently this method has been extended to determine constituent masses for diquarks with all possible flavors (see Tables~\ref{LightDiquarkMassesTable}--\ref{BottomDiquarkMassesTable} for a summary of QCD sum rule diquark mass determinations). Generally speaking, constituent diquark masses determined via QCD sum rules are in good agreement with the diquark mass parameters used in compact hadron models, providing QCD support for the conclusions of those models. 

To date, crucial NLO perturbative contributions have been incorporated into QCD sum rule calculations of diquarks with all possible flavors, except for the $[bc]$ diquark.\footnote{The $[bc]$ notation is used in in this work to denote the antisymmetric color anti-triplet configuration.} The aim of this work is to close this gap.
Ref.~\cite{Tang:2012np} studied $J^{P}\in\{0^+,1^+\}$ $[bc]$ diquarks with QCD sum rules, including  leading order (LO) perturbative effects and truncating the operator product expansion (OPE) at dimension-four. We build upon this work by incorporating NLO perturbative corrections, including dimension-six terms in the OPE, and by extending our analysis to diquarks with negative parity. The presence of three distinct scales ({\it i.e.}, $m_c$, $m_b$, and $q^2$) means that the loop integrals required for a full NLO perturbative calculation are formidable. Existing analytic expressions~\cite{Scharf:1993ds,Martin:2003qz} 
for the resulting multi-scale master integrals are not known to the required order in $\epsilon$ and are insufficient for a NLO calculation requiring renormalization. Therefore we adopt a numerical approach wherein OPE Wilson coefficients are expressed in terms of analytically calculated master integral coefficients, and the master integrals themselves are calculated numerically using pySecDec~\cite{Borowka:2017idc,Borowka:2018goh,Heinrich:2021dbf,Heinrich:2023til}.\footnote{pySecDec was also used in Ref.~\cite{Ray:2022fcl} to compute NLO perturbative contributions for a QCD sum rule analysis of a $\bar{u}\bar{d}ud$ tetraquark.} The numerical contour integration in the complex $q^2$ plane used in Ref.~\cite{Esau:2019hqw} to construct Laplace sum rules is naturally complemented by these numerical loop integration techniques. Using this innovative numerical approach, we surmount the challenge of the multiple scales intrinsic to $[bc]$ diquarks and perform QCD sum rule analyses to determine their masses. 

The remainder of the paper is organized as follows. In Section~\ref{II} we define $[bc]$ diquark currents and their correlation functions with $J^{P}\in\{0^\pm\,,1^\pm\}$, and then perform the QCD calculation in the OPE including NLO perturbative contributions and operators up to dimension-six. In Section~\ref{III} we discuss our numerical construction of the QCD Laplace sum rules, their analysis, and the resulting $[bc]$ diquark masses. In Section~\ref{IV} we conclude by comparing our $[bc]$ diquark mass predictions to the mass parameters used in diquark models of $T_{bc}$ tetraquarks, and we also comment on the hierarchy of diquark masses for all possible flavor combinations as determined by QCD sum rules. 

\section{The Correlator}\label{II}

The $[bc]$ color antitriplet diquark current is given by~\cite{Dosch1989,Jamin:1989hh}
\begin{equation}
  j_{\alpha} = \epsilon_{\alpha\beta\gamma}b_{\beta}^{T} C G c_{\gamma},
  \label{Current}
\end{equation}
where
\begin{equation}
  G \to \{\gamma^{5}, I, \gamma_{\mu}, \gamma_{\mu}\gamma^{5} \}\label{OOO}
\end{equation}
for the scalar ($0^+$), pseudoscalar ($0^-$), axial vector ($1^+$) and vector ($1^-$) channels respectively. The adjoint of $j_{\alpha}$ is given by
\begin{equation}
  j_{\alpha}^{\dag} = \epsilon_{\alpha\beta\gamma} \overline{c}_{\gamma}\widetilde{G} C \overline{b}_{\beta}^{T}, \label{CurrentDag}
\end{equation}
with
\begin{equation}
  \widetilde{G} \to \{\gamma^{5}, -I, -\gamma_{\mu}, \gamma^{5}\gamma_{\mu} \}.\label{OOODag}
\end{equation}
Here $C$ denotes the charge conjugation operator, $\epsilon_{\alpha\beta\gamma}$ is a Levi-Civita symbol in quark color space, $I$ is the identity matrix and $\{b, c\}$ are bottom and charm quark fields respectively.

Using equations~(\ref{Current}) to~(\ref{OOODag}), in the appropriate combinations we consider the diquark correlator
\begin{equation}
  \Pi(q^2) = P
  \int d^{D}\!x \;e^{iq\cdot x}
  \vev{\tau[\, j_{\alpha}(x)\; S_{\alpha\omega}(x,0)\; j_{\omega}^{\dag}(0)]},
  \label{Correlator}
\end{equation}
where $P$ is the projection operator used to extract the $(J=0)$ scalar or $(J=1)$ vector portion of the correlation function:
\begin{equation}
  P \in \Big\{ 1 , \;\; \frac{i}{D-1}
  \left(\frac{q_{\mu}q_{\nu}}{q^2}-g_{\mu\nu}\right) \Big\}
  \label{projection}
\end{equation}
for $J \in \{ 0,1 \}$ respectively. Here $D$ is the spacetime dimension. $S_{\alpha\omega}(x,0)$ is a path-ordered exponential, or Schwinger string, given by
\begin{equation} \label{SchwingerString}
S_{\alpha\omega}(x,0) = \hat{\mathcal{P}}\; \text{exp} \bigg[ i g_{s} \frac{\lambda^{a}_{\alpha\omega}}{2} \int_{0}^{x} dz^{\mu} A_{\mu}^{a}(z) \bigg],
\end{equation}
where $\hat{\mathcal{P}}$ is the path-ordering operator. 
The Schwinger string ensures that gauge-invariant information can be extracted from the gauge-dependent current \eqref{Current}~\cite{Dosch1989, Jamin:1989hh}.   
The explicit cancellation of the gauge parameter has been demonstrated for perturbative contributions through NLO~\cite{Kleiv:2013dta}, and in Landau gauge the NLO contributions from the Schwinger string vanish~\cite{Dosch1989, Jamin:1989hh}, allowing for the replacement $S_{\alpha\omega}(x,0) \to \delta^{\alpha\omega}$. For the non-perturbative QCD condensate contributions, gauge invariance of the correlator~\eqref{Correlator} ensures that OPE coefficients obtained using fixed-point gauge are equivalent to those obtained using other methods~\cite{Bagan:1992tg}. Furthermore, as noted in Refs.~\cite{Dosch1989, Jamin:1989hh}, the Schwinger string does not contribute to the condensate terms in fixed-point gauge, again implying $S_{\alpha\omega}(x,0) \to \delta^{\alpha\omega}$.
Consequently, by evaluating perturbative contributions in Landau gauge and condensate contributions using fixed-point gauge methods, the correlator~(\ref{Correlator}) may be simplified by setting $S_{\alpha\omega}(x,0) \to \delta^{\alpha\omega}$ as in Ref.~\cite{Kleiv:2013dta}.

We evaluate the correlator~(\ref{Correlator}) within the OPE to NLO in perturbation theory, including non-perturbative corrections proportional to the dimension-four (4d) and dimension-six (6d) gluon condensates and the 6d quark condensate. 
Each non-perturbative contribution consists of a LO perturbatively computed Wilson coefficient multiplied by the corresponding QCD condensate. The Wilson coefficients have been calculated in Mathematica using FeynCalc~\cite{Mertig:1990an,Shtabovenko:2016sxi,Shtabovenko:2020gxv,Shtabovenko:2023idz} combined with in-house code that also draws upon HypExp  \cite{Huber:2005yg,Huber:2007dx}. 
The 4d and 6d gluon condensates and the 6d quark condensate are defined, respectively, by
\begin{equation} \label{fourDcond}
\big\langle \alpha G^{2} \big\rangle = \alpha_{s} \big\langle \! \colon \! G_{\omega\phi}^{a} G_{\omega\phi}^{a} \! \colon\! \big\rangle,
\end{equation}
\begin{equation} \label{sixDcond}
\begin{aligned}
\big\langle g^{3}G^{3}\big\rangle = g_{s}^{3} f^{abc} \big\langle \! \colon \! G^{a}_{\omega\zeta} \; G^{b}_{\zeta\rho} G^{c}_{\rho\omega} \! \colon\! \big\rangle,
\end{aligned}
\end{equation}
\begin{equation} \label{sixDQcond}
\big\langle J^2 \big\rangle
= \frac{D}{6} \, \kappa \, g_s^4 \big\langle \overline{q}q \big\rangle^2,
\end{equation}
where $\kappa$ in~(\ref{sixDQcond}) quantifies deviation from vacuum 
saturation. As in~\cite{Palameta:2017ols,Palameta:2018yce}, we set $\kappa=2$ for the remainder of this calculation, 
\eg\ see~\cite{narisonbook:2004} and references contained therein.

\begin{figure}[hbt]
\centering
\begin{tabular}{ccccc}
\includegraphics[width=30mm]{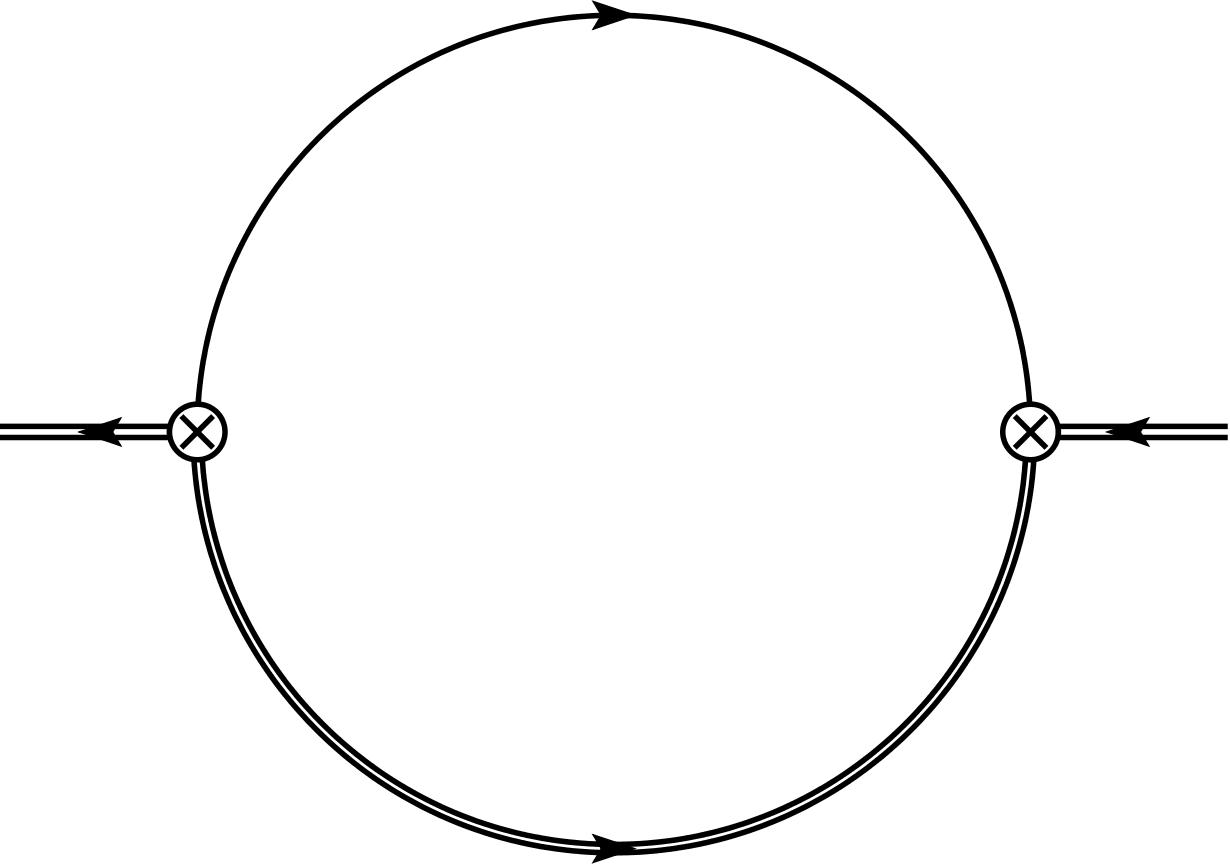} & 
\includegraphics[width=30mm]{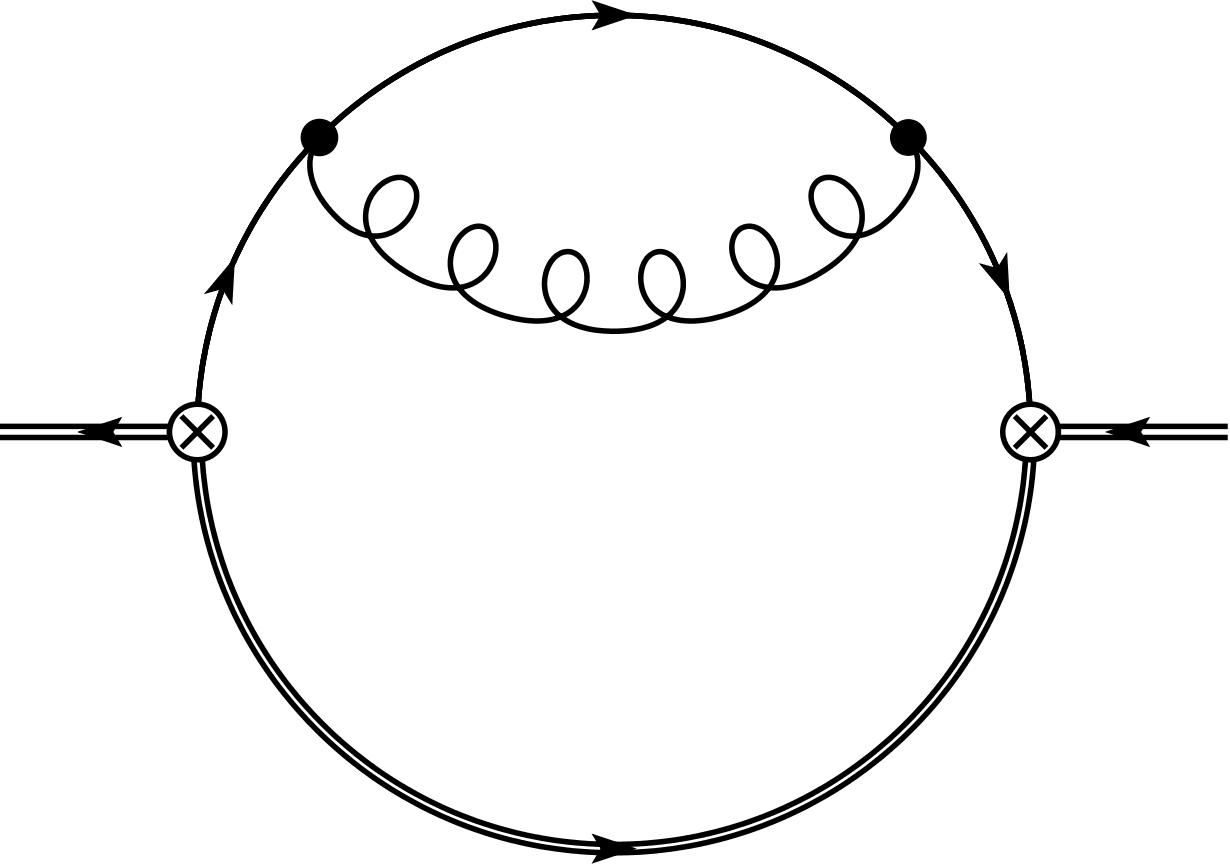} & 
\includegraphics[width=30mm]{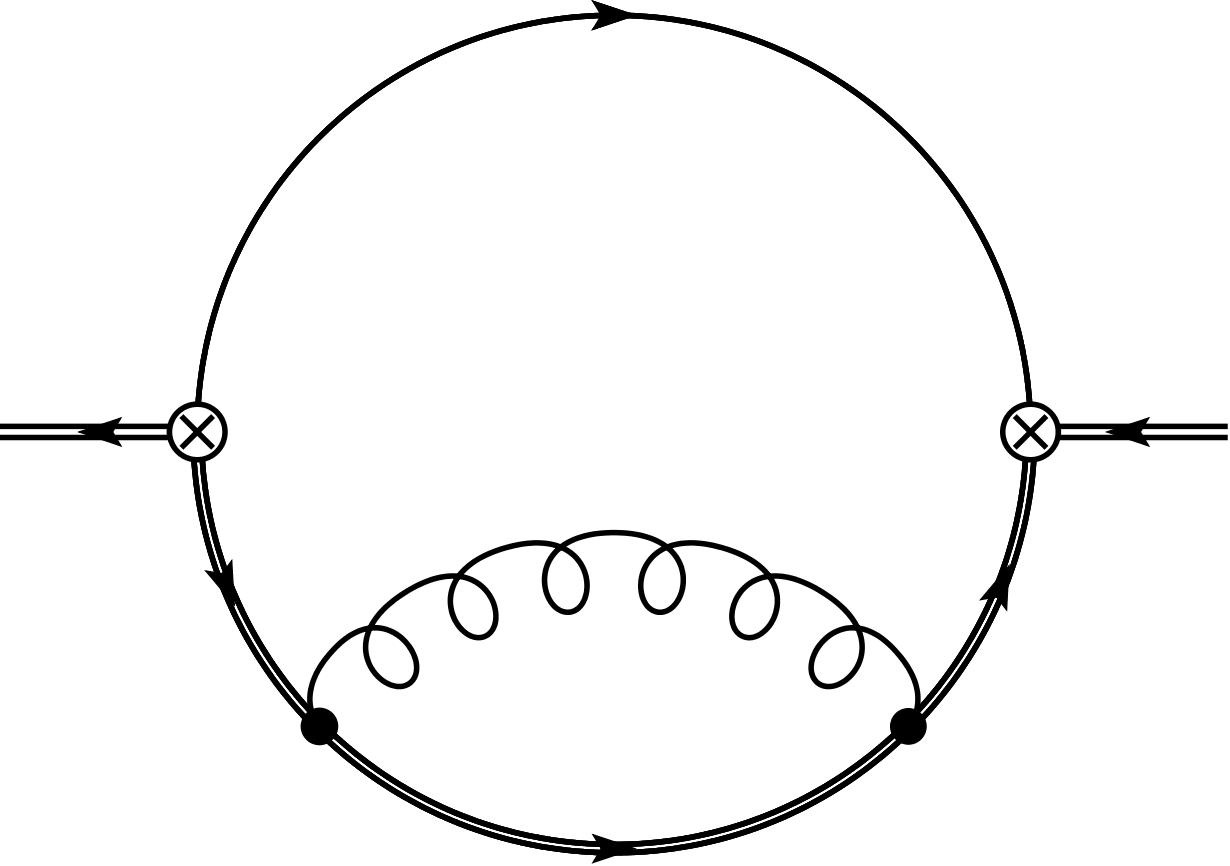} &
\includegraphics[width=30mm]{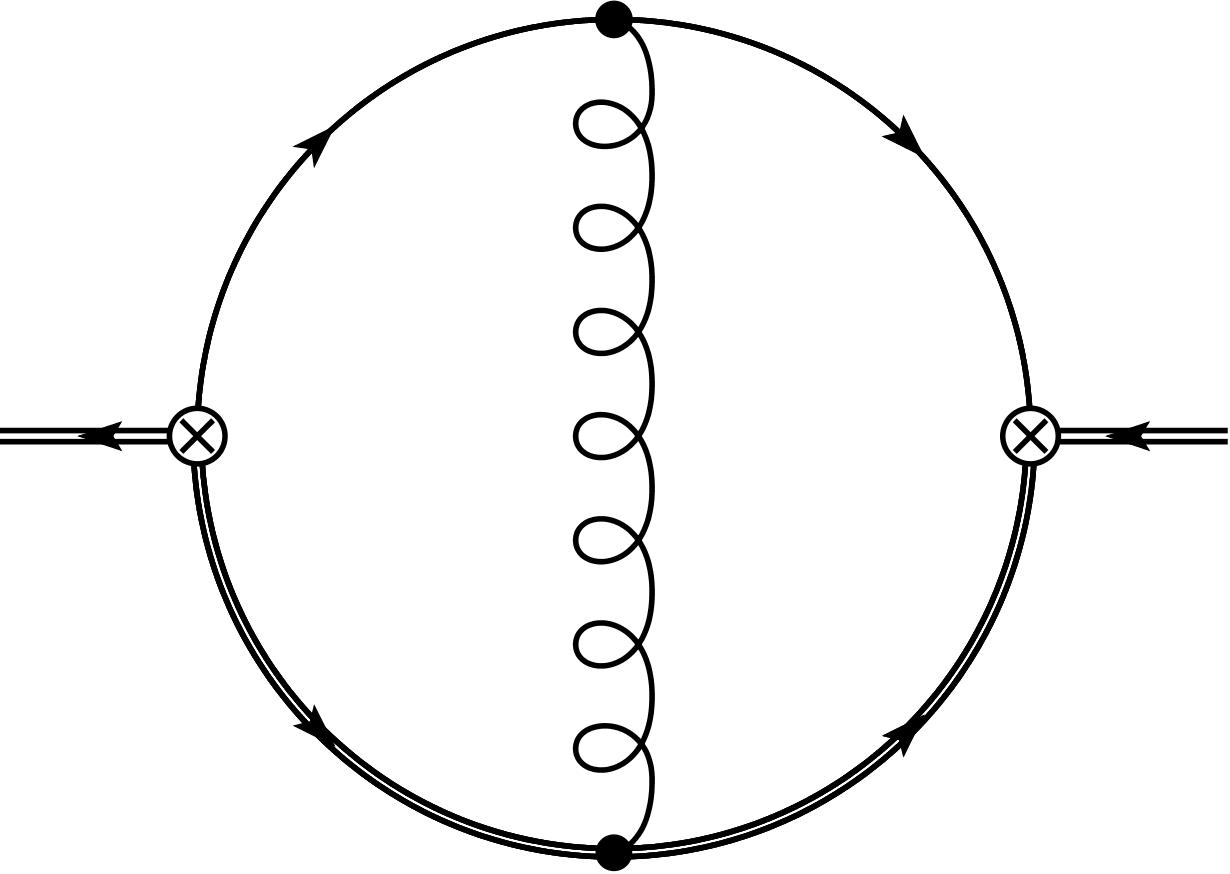} &
\includegraphics[width=30mm]{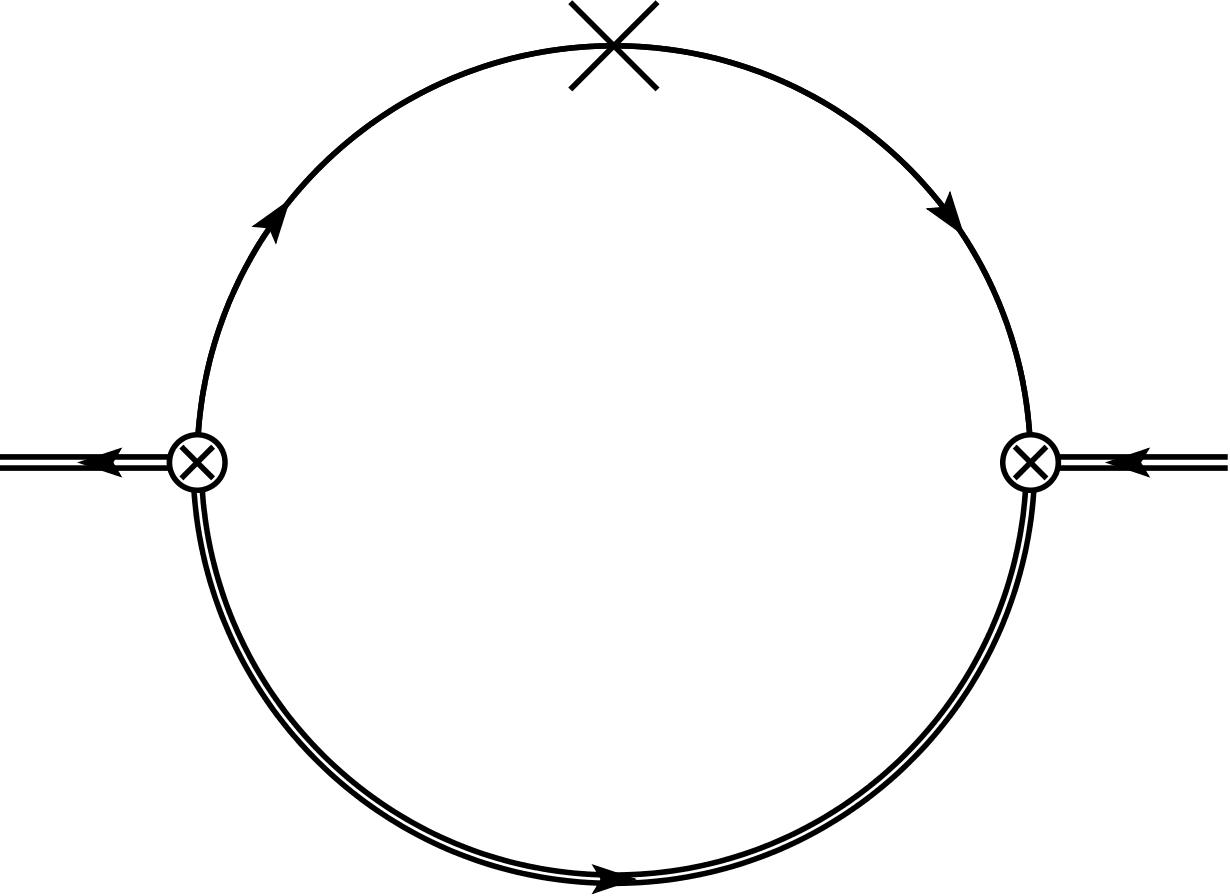} \\
Diagram \rom{1} & Diagram \rom{2} & Diagram \rom{3} & Diagram \rom{4} & Diagram C2 \\[15pt]
\includegraphics[width=30mm]{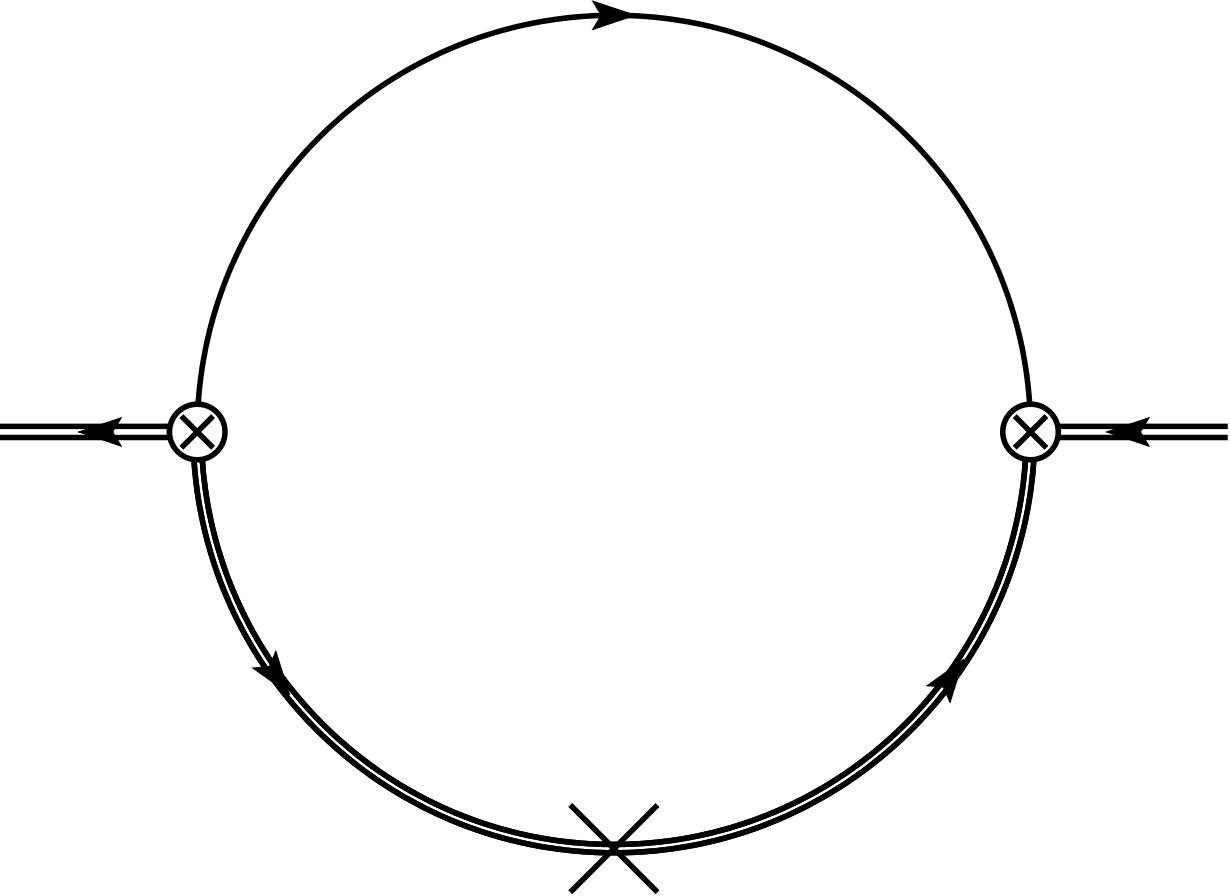} & 
\begin{overpic}[width=30mm,tics=10]{11.jpg}
 \put (25,31) {$(Z_d^2 - 1)$}
\end{overpic} & 
\includegraphics[width=30mm]{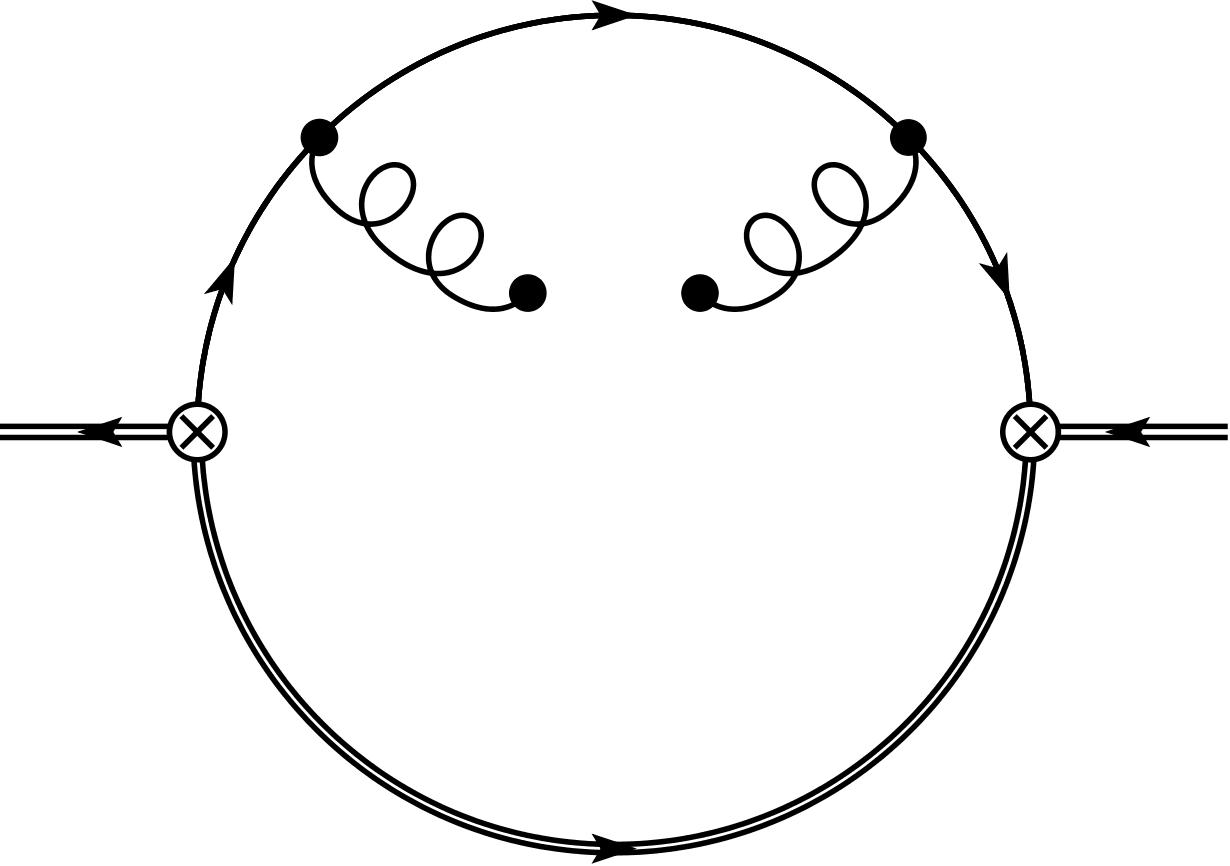} & 
\includegraphics[width=30mm]{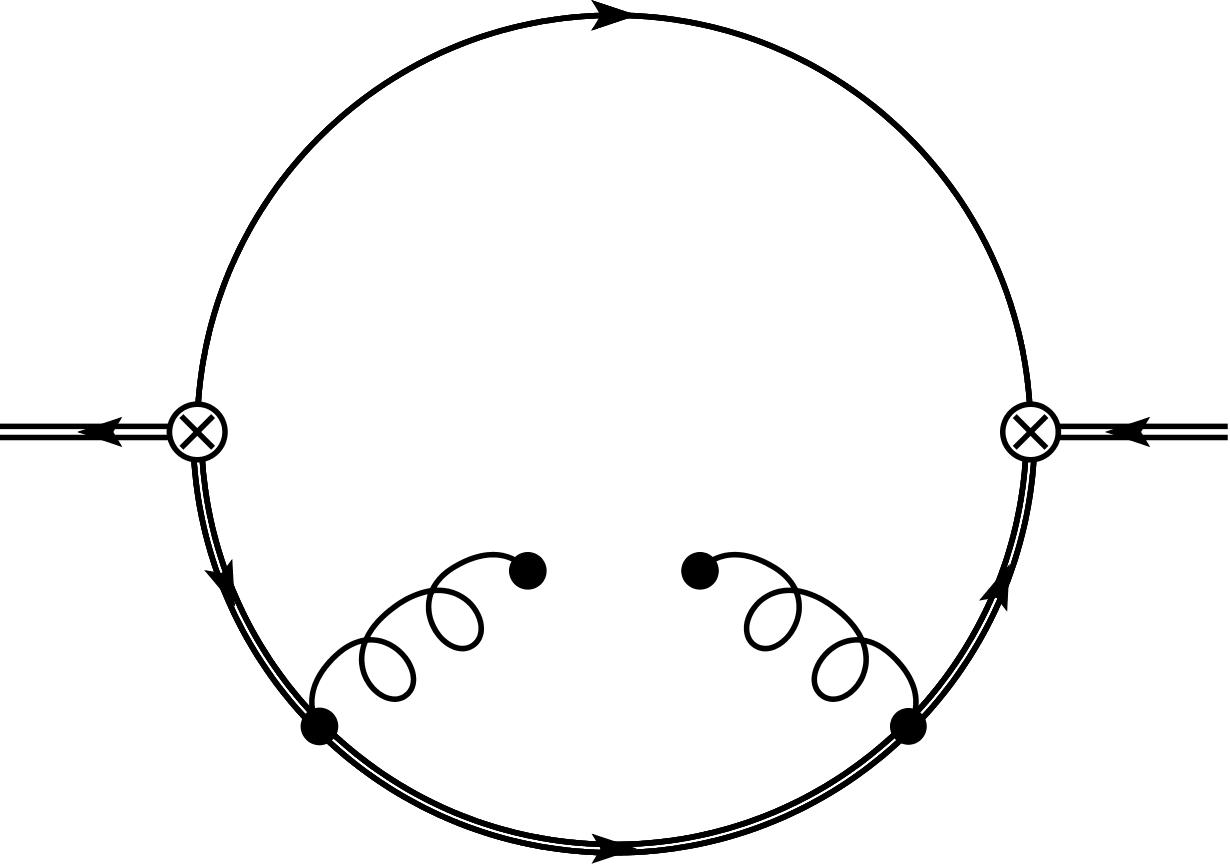} & 
\includegraphics[width=30mm]{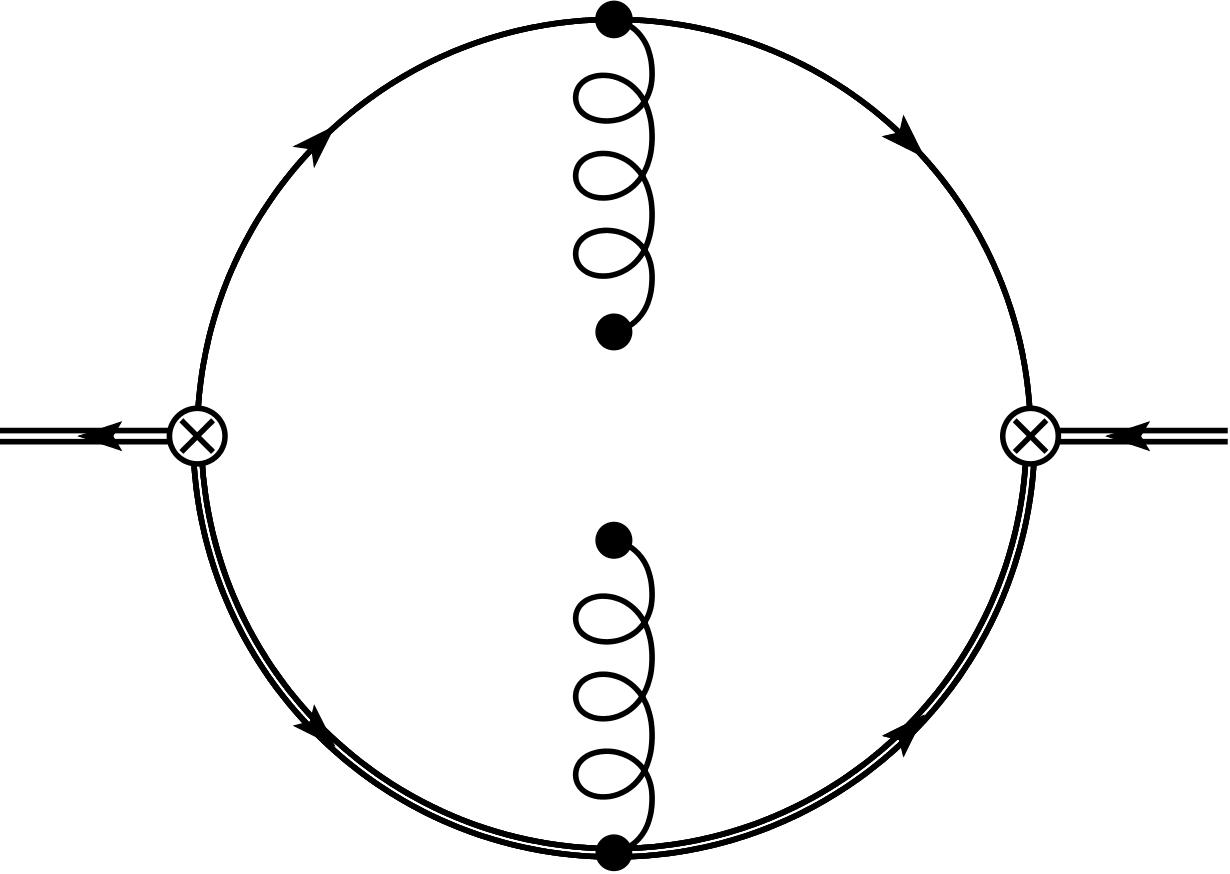} \\
Diagram C3 & Diagram C4 & Diagram \rom{5} & Diagram \rom{6} & Diagram \rom{7} \\[15pt]
\includegraphics[width=30mm]{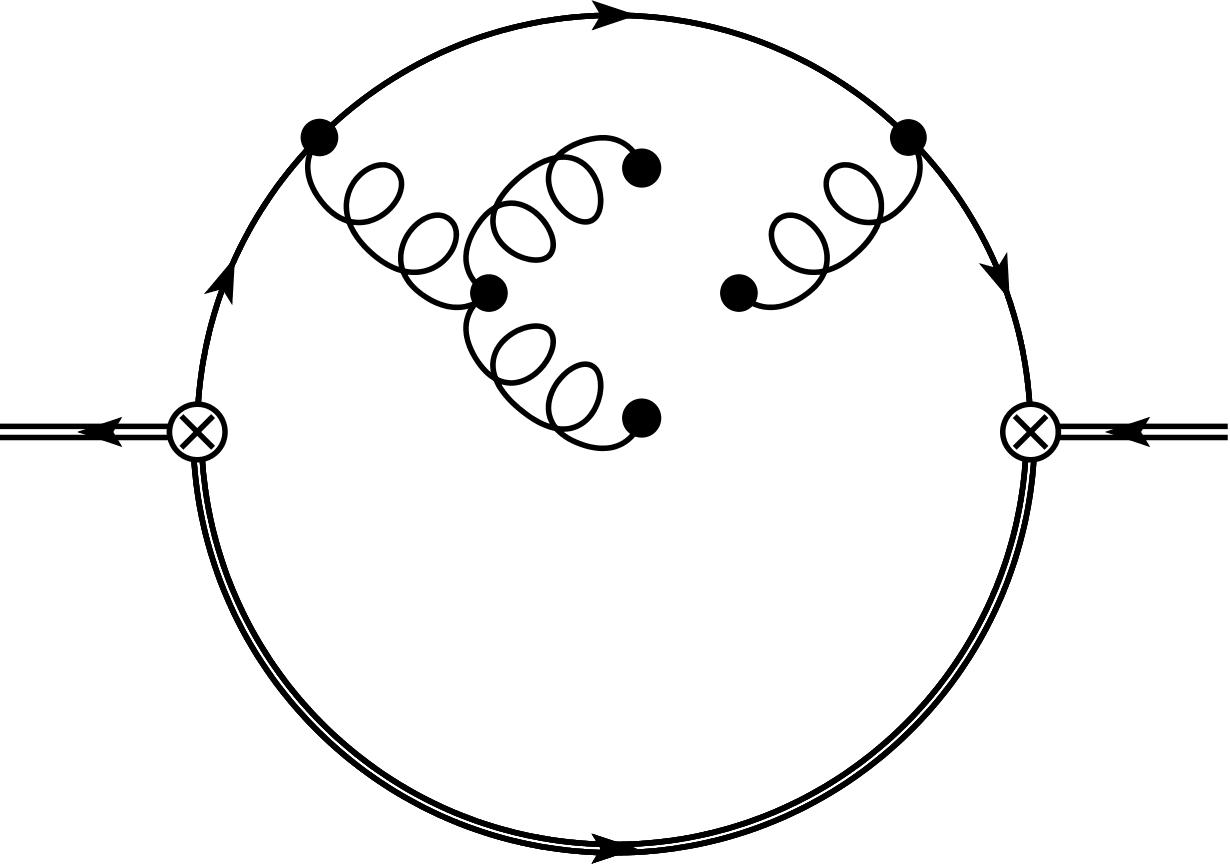} & 
\includegraphics[width=30mm]{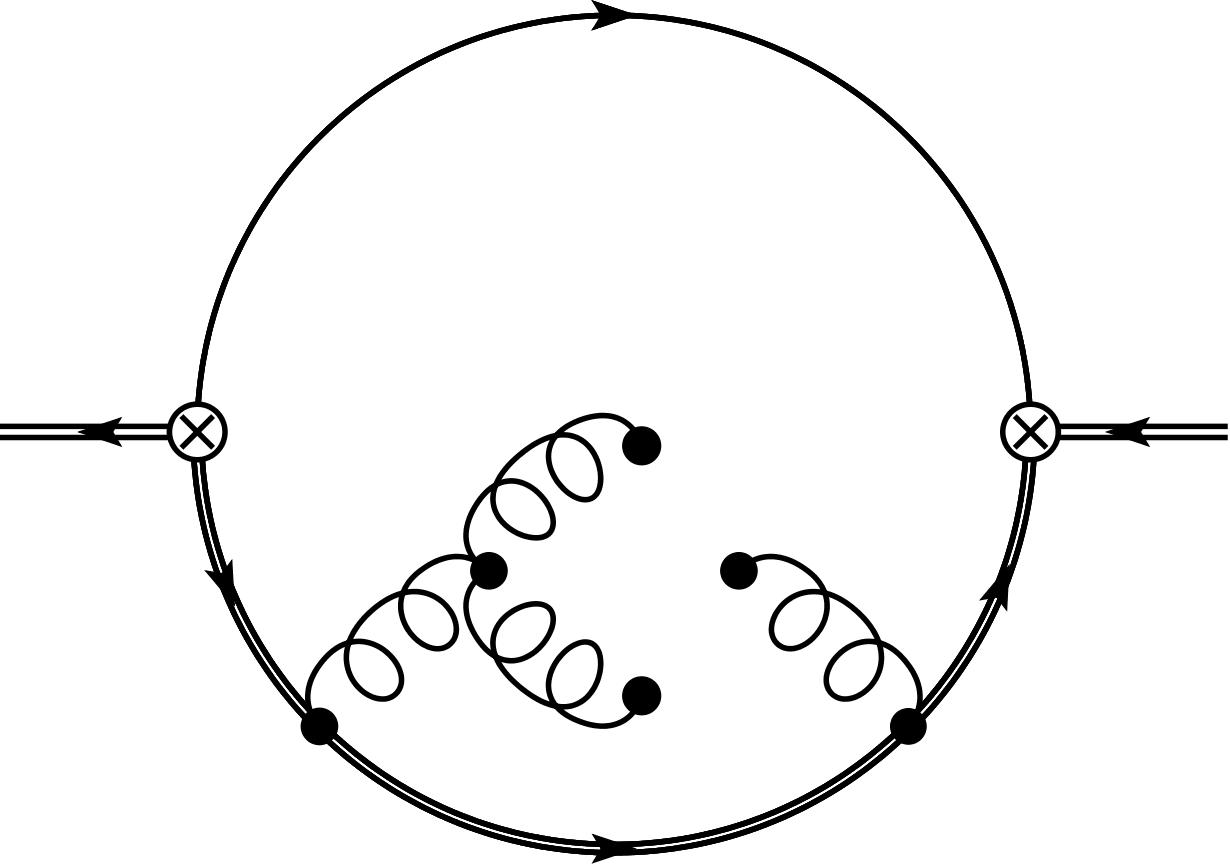} & 
\includegraphics[width=30mm]{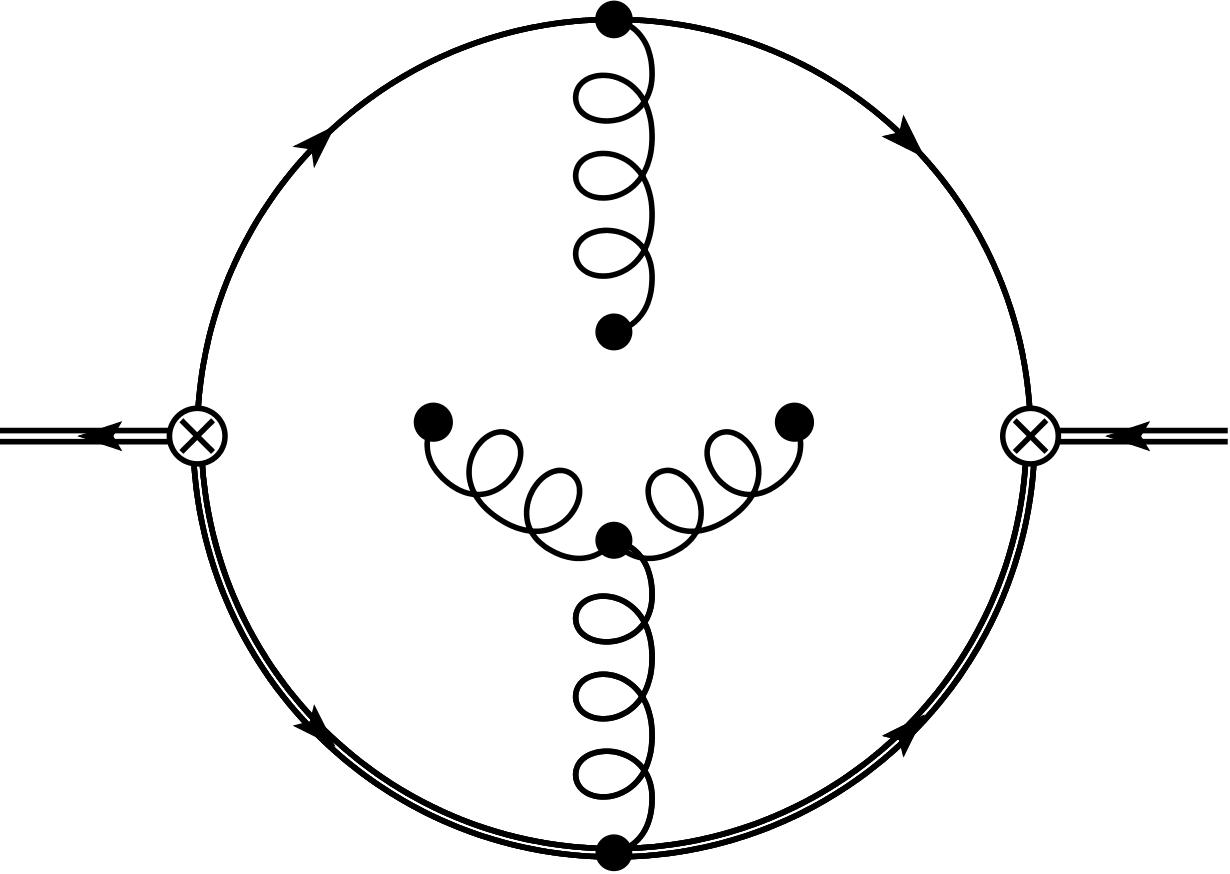} &
\includegraphics[width=30mm]{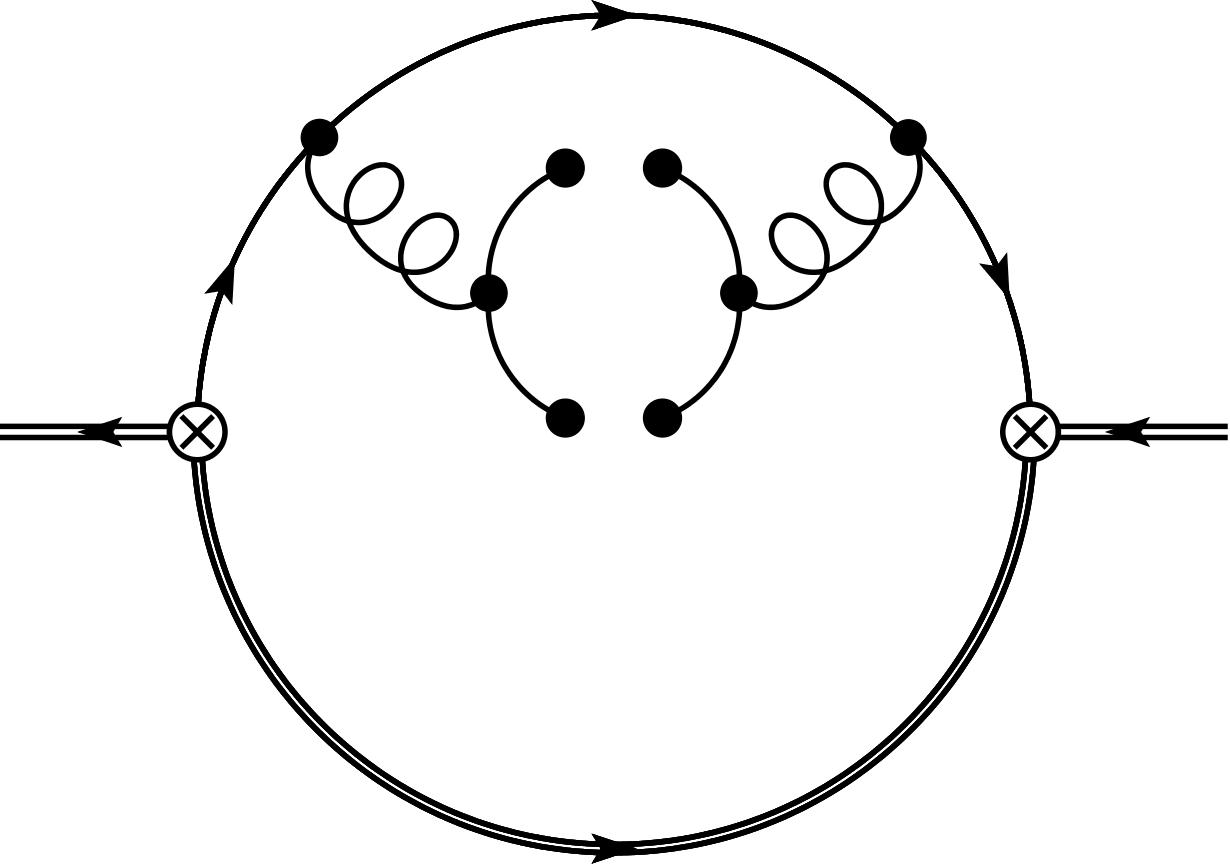} &
\includegraphics[width=30mm]{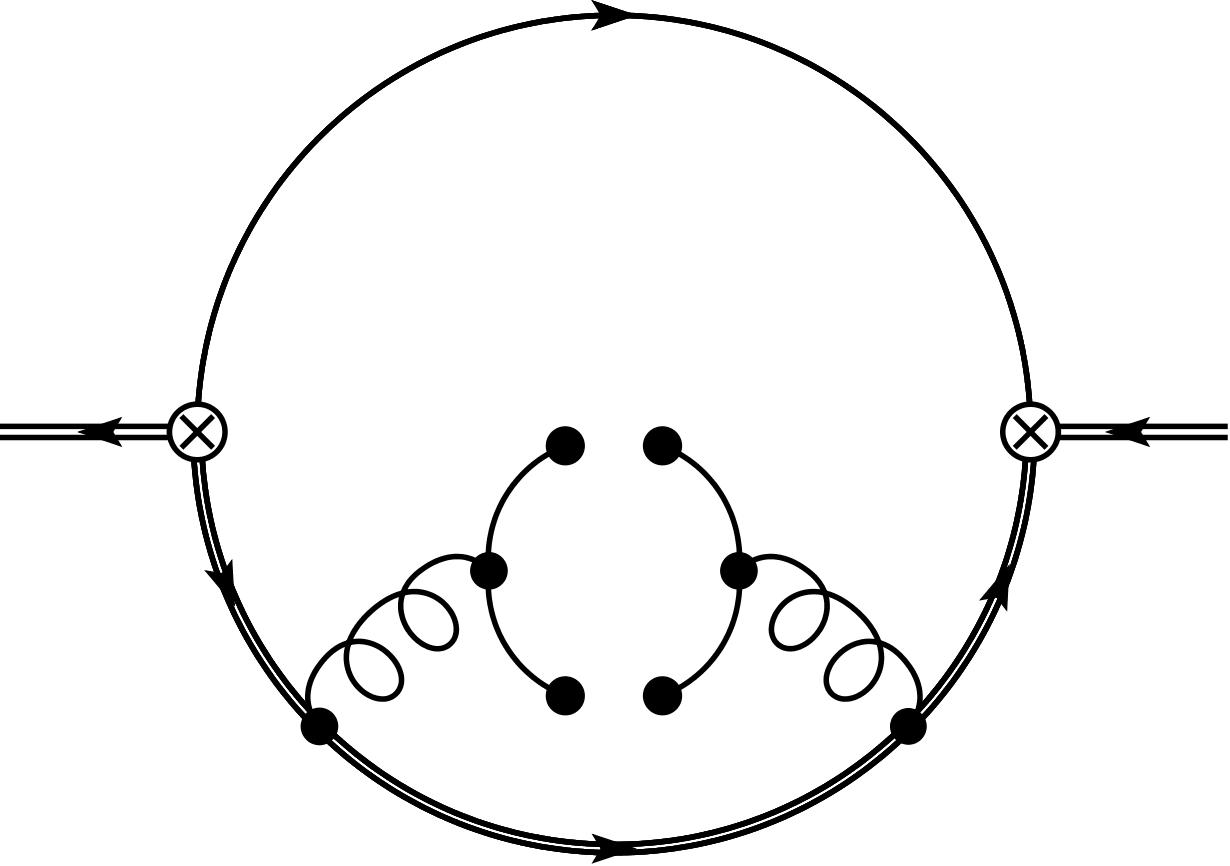} 
\\
Diagram \rom{8} & Diagram \rom{9} & Diagram \rom{10} & Diagram \rom{11} & Diagram \rom{12} \\[15pt]
\includegraphics[width=30mm]{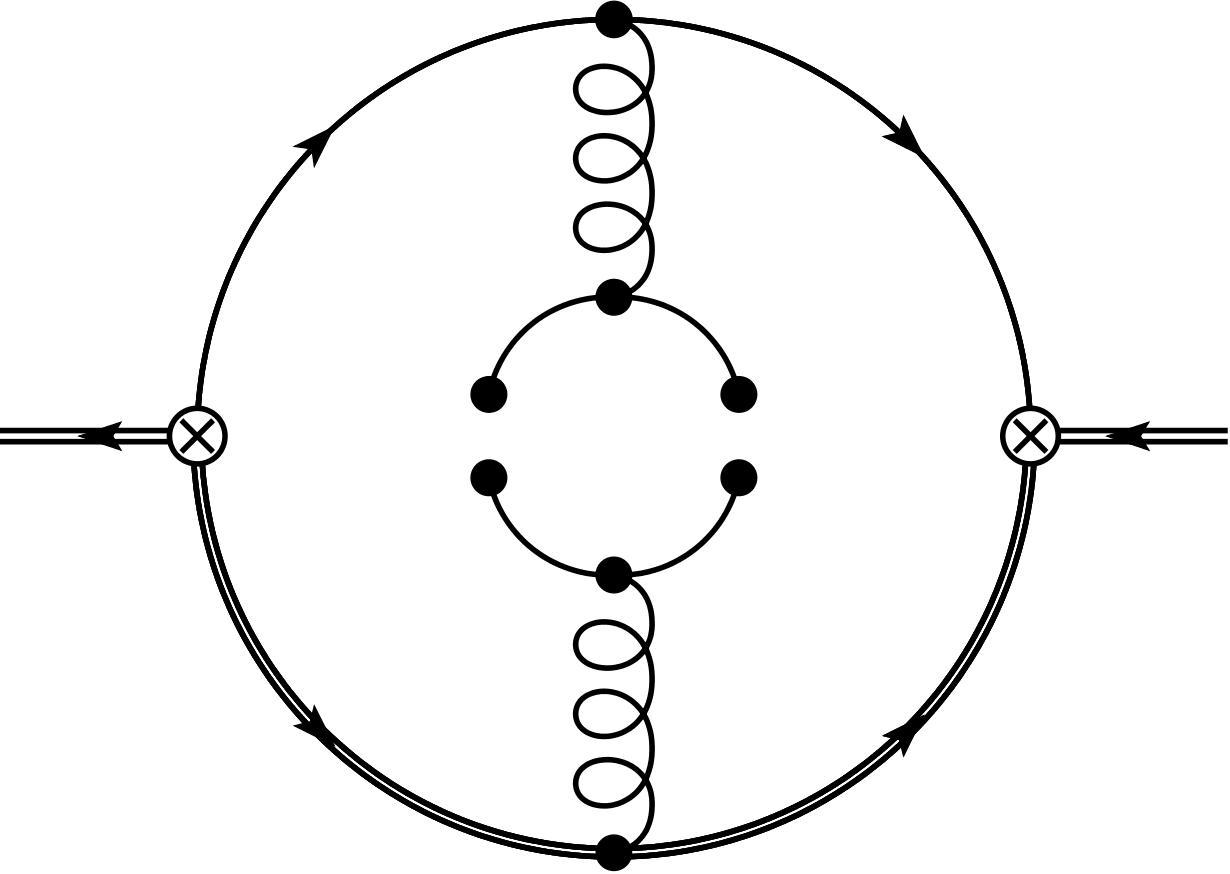} & 
\includegraphics[width=30mm]{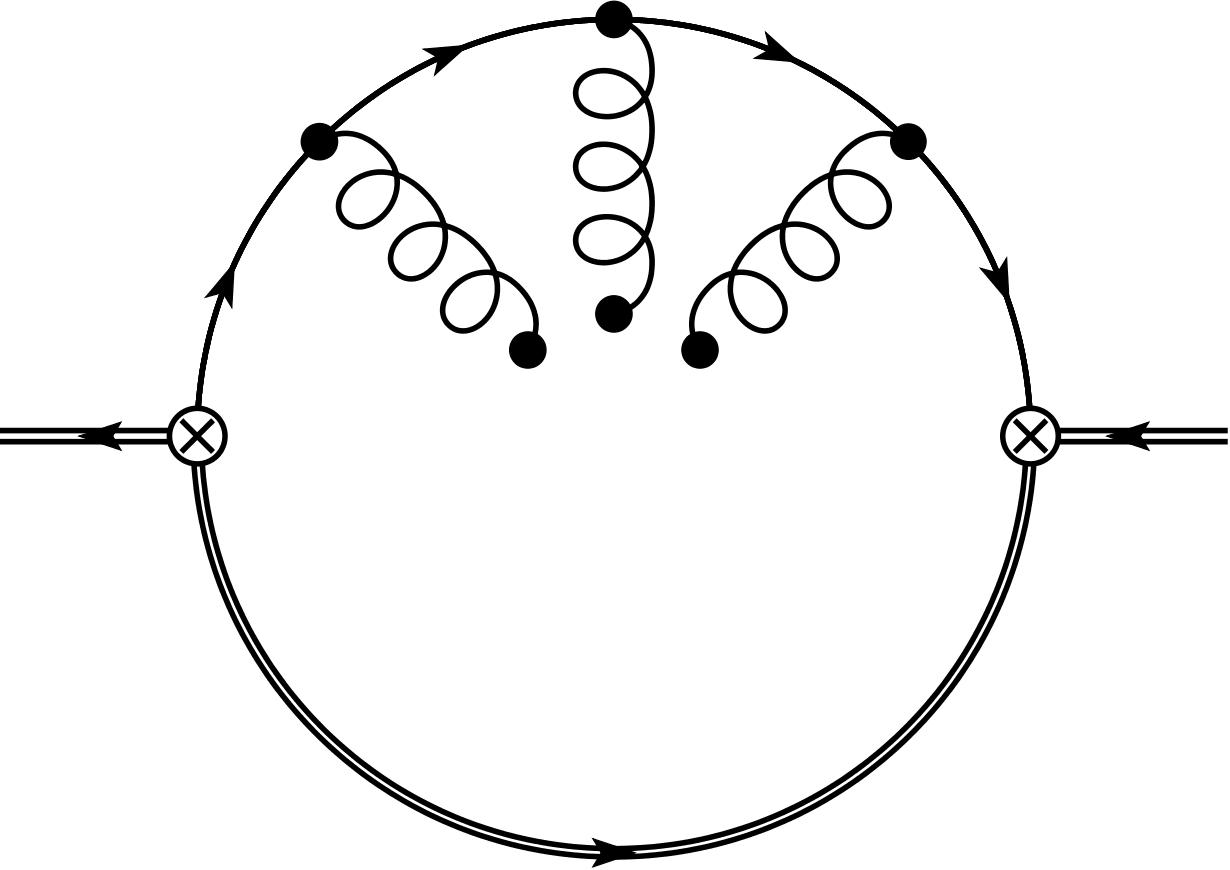} & 
\includegraphics[width=30mm]{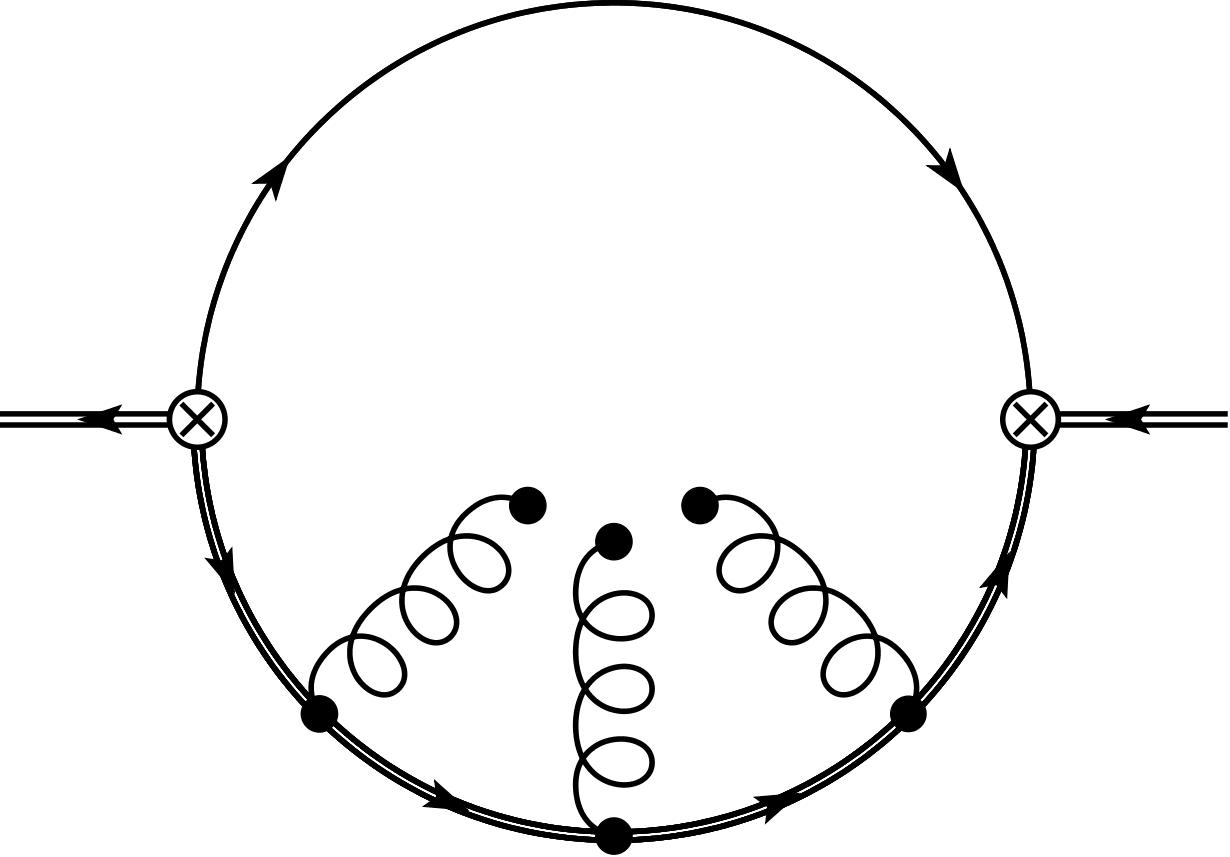} &
\includegraphics[width=30mm]{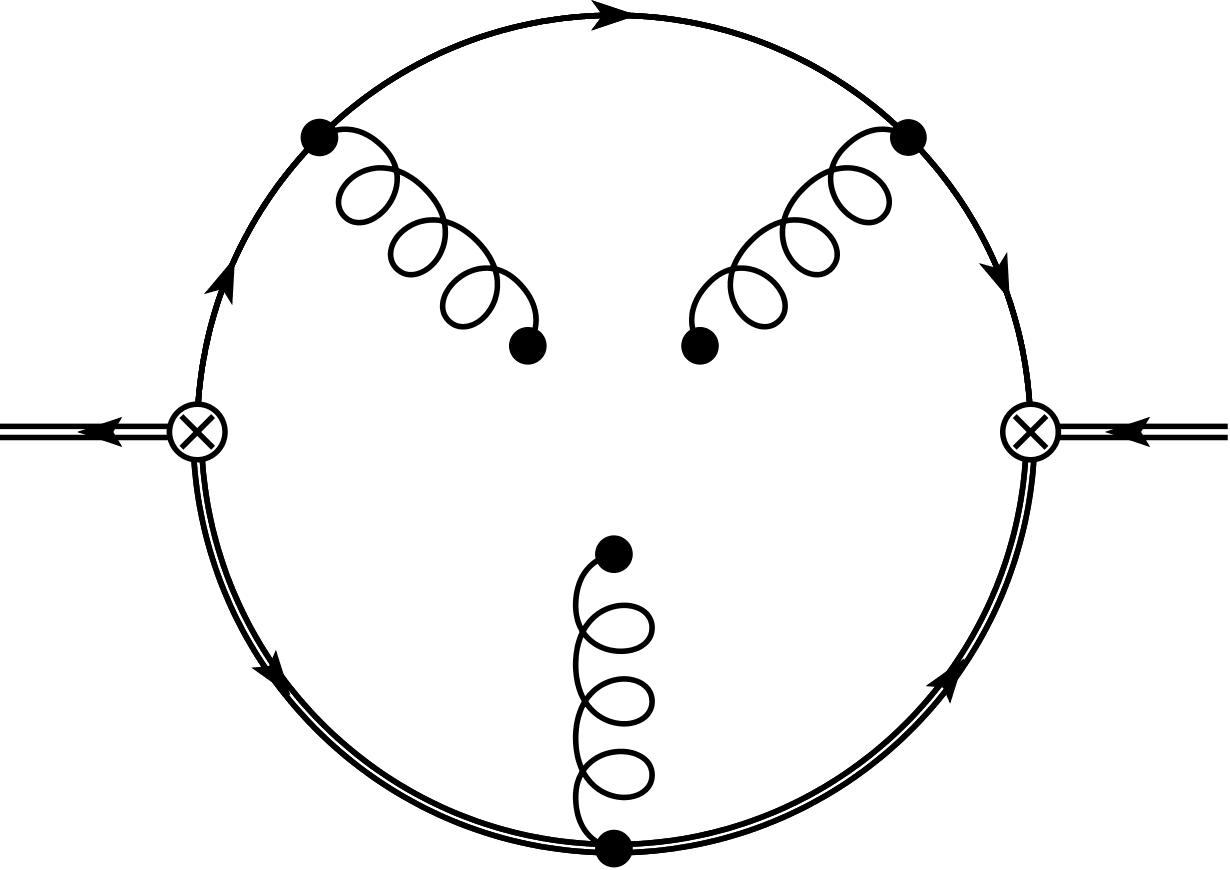} &
\includegraphics[width=30mm]{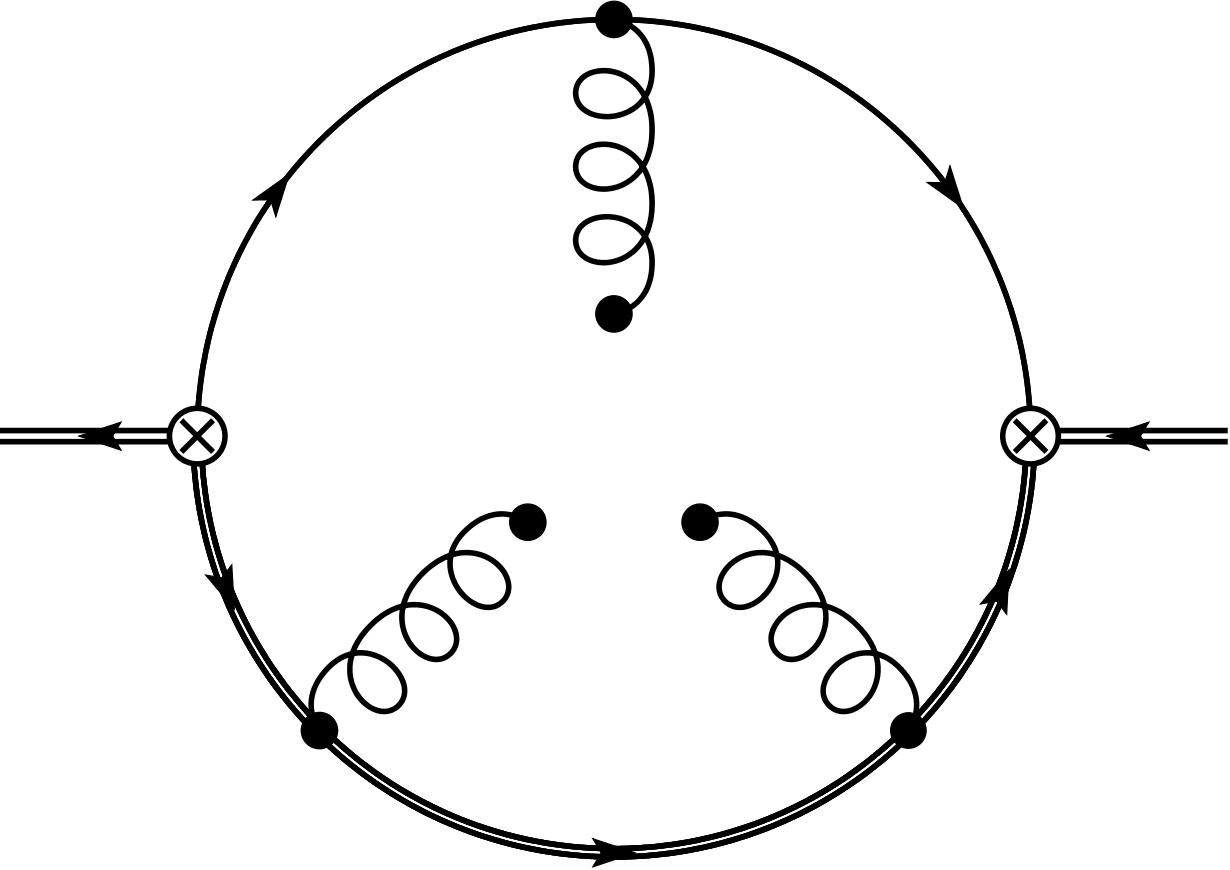} 
\\
Diagram \rom{13} & Diagram \rom{14} & Diagram \rom{15} & Diagram \rom{16} & Diagram \rom{17} \\[15pt]
\end{tabular}
\caption{Perturbative diagrams, counter-term diagrams, and non-perturbative diagrams. Here the single line corresponds to a $c$-quark and the double line corresponds to a $b$-quark. Feynman diagrams were created using JaxoDraw \cite{Binosi:2003yf}.} 
\label{fig01}
\end{figure}

The diagrams computed in the simplification of~(\ref{Correlator}) are given in Figure~\ref{fig01}. Diagram \rom{1} corresponds to the LO perturbative contribution, diagrams \rom{2} to \rom{4} are the NLO perturbative contribution and diagrams C2 to C4 are the counter-term diagrams that serve to renormalize diagrams \rom{2} to \rom{4}. Diagrams \rom{5} to \rom{7} correspond to the 4d gluon condensate contributions, diagrams \rom{8} to \rom{10} are 6d gluon condensate contributions and diagrams \rom{11} to \rom{13} are 6d quark condensate contributions. Diagrams \rom{14} to \rom{17} are additional 6d quark contributions that account for topologies involving three interaction vertices that contribute at this order. As noted earlier, Wilson coefficients are calculated in the Landau gauge. Divergent integrals are handled using dimensional regularization in $D=4+2\epsilon$ dimensions at $\overline{\text{MS}}$ renormalization scale $\mu$. We use a dimensionally regularized $\gamma^5$ satisfying $(\gamma^5)^2=1$ and $\{\gamma^{\mu},\,\gamma^5\}=0$~\cite{ChanowitzFurmanHinchliffe1979}. In Diagram C4 of Fig.~\ref{fig01} the one-loop multiplicative renormalization factor for the diquark currents is given by
\begin{equation}
Z_d^2-1 = \begin{cases}
    \frac{\alpha_s}{\pi}\frac{1}{\epsilon} \,, \quad \text{in the scalar $(0^+)$ and pseudoscalar $(0^-)$ channels,}
    \\
    0 \,,  \quad \text{ in the axial vector $(1^+)$ and vector $(1^-)$ channels,}
\end{cases}    
\end{equation}
as calculated in Ref.~\cite{Kleiv:2010qk}.

The recurrence relations of Refs.~\cite{Tarasov1996,Tarasov1997} are implemented via the Mathematica package Tarcer~\cite{MertigScharf1998} resulting in expressions phrased in terms of master integrals. Our in-house Mathematica code contains implementations of the  master integrals given in Refs.~\cite{BoosDavydychev1991,BroadhurstFleischerTarasov1993} which we use for benchmarking purposes. In particular, the non-trivial Tarcer-notation master integrals  that arise are
 \begin{gather}
 \text{TJI}[D, q^2, \{\{n_1, m_c\}, \{n_2, m_b\}, \{1, 0\}\}], \label{eq:Jints}
 \quad
 \left(n_1\,,n_2\right) \in \{ (1,1)\,, \, (1,2)\,, \, (2,1) \},
 \\
 \text{TBI}[D, q^2, \{\{1, m_c\}, \{1, m_b\}\}]. \label{eq:Bint}
 \end{gather}
Renormalization of the NLO perturbative contributions requires the $\cal{O}(\epsilon)$ terms in the $\epsilon$-expansions of the integrals in Eqs.~\eqref{eq:Jints} and \eqref{eq:Bint}. 
Collectively, the existing analytic expressions for these integrals~\cite{Scharf:1993ds,Martin:2003qz}  are not known to the required order in $\epsilon$ and  we have therefore evaluated them numerically to $\cal{O}(\epsilon)$ using pySecDec \cite{Borowka:2017idc,Borowka:2018goh,Heinrich:2021dbf,Heinrich:2023til} (pySecDec also uses CUBA \cite{Hahn:2004fe,Hahn:2014fua}, FORM \cite{Vermaseren:2000nd,Kuipers:2013pba,Ruijl:2017dtg}, and 
GSL \cite{galassi2018scientific}). 
Details on our pySecDec implementation are outlined below.

We proceed numerically with the OPE computation of the correlation function, which can be written as
\begin{equation} \label{decomposedCorrelator}
\Pi(q^2)=
\sum_{\text{i}=\text{\rom{1}}}^{\text{\rom{17}}}\Pi^{\text{(i)}}(q^2)
\end{equation}
where the superscript in~(\ref{decomposedCorrelator}) corresponds to the labels of the diagrams in Figure~\ref{fig01}. Note that all diagrams are summed over, including the counter-term diagrams, which results in a removal of all non-local divergences. The results are then expanded in $\epsilon$ using pySecDec and we proceed with the analysis. 
Due to the presence of multiple scales ($m_c$, $m_b$, and $q^2$) and our numerical approach, expressions for individual Feynman diagrams are very large and unwieldy, for the sake of brevity they are not included here.

Renormalization-group improvement requires that the strong coupling and quark mass be replaced by their corresponding running quantities evaluated at renormalization scale $\mu$~\cite{Narison:1981ts}. At one-loop in the $\overline{\text{MS}}$ renormalization scheme, for $[bc]$ diquarks which naturally have a renormalization scale in the $n_f=4$ region between the charm and bottom thresholds, we have
\begin{gather}
  \alpha_s(\mu) = \frac{\alpha_s(M_{\tau})}{1 + \frac{25 \alpha_s(M_{\tau})}%
  {12\pi}\log\!{\Big(\frac{\mu^2}{M_{\tau}^2}\Big)}},
  \\ 
 m_{c}(\mu) = \overline{m}_{c}\bigg(\frac{\alpha_s(\mu)}
  {\alpha_s(\overline{m}_{c})}\bigg)^{12/25},
\\
 m_{b}(\mu) = \overline{m}_{c}\bigg(\frac{\alpha_s(\mu)}
  {\alpha_s(\overline{m}_{b})}\bigg)^{12/25},
\end{gather}
where \cite{ParticleDataGroup:2024cfk}\footnote{The running strong coupling at $M_\tau$ is not specified in the most recent PDG listing~\cite{ParticleDataGroup:2026aaa}. However, starting from the currently quoted value of $\alpha_s(M_Z)$ and evolving it from the $n_f=5$ region to the $n_f=4$ region across the bottom-quark mass threshold, we obtain a value of $\alpha_s(M_\tau)$ that is consistent with the one used here. A transition relation of $\alpha_s$ at the bottom mass threshold is also taken into consideration as in Ref.~\cite{Steele:1998qls}.
}
\begin{gather}
 \alpha_s\left(M_\tau\right)=  0.314\,,~M_\tau= 1776.93 \,{\rm MeV} \label{a1}, \\
 \overline{m}_c=m_c\left(\overline{m}_c\right)=\left(1.2730 \pm 0.0046 \right),
 \label{a3} 
 \\
  \overline{m}_b=m_b\left(\overline{m}_b\right)=\left(4.183\pm 0.007  \right) \, {\rm GeV}\,.
\label{a4}
\end{gather}
For $[bc]$ diquarks our central value analysis follows the $bc$ hybrid meson sum-rule calculation of Ref.~\cite{Chen:2013eha} by choosing an intermediate $\overline{\rm MS}$ renormalization scale $\bar\mu$ given by
\begin{equation}
\bar\mu=\frac{\overline{m}_c\left(m_c\right)+m_b\left(\overline{m}_b\right)}{2}   =  2.728  \, {\rm GeV}\,,
\label{tom_renorm_scale}
\end{equation}
along with the corresponding values 
\begin{gather}
\tilde\alpha_s=\alpha_s\left(\bar\mu\right) =0.266434\,,
\label{tom_tilde_alpha}
\\
\tilde m_c=m_c\left(\bar\mu\right)=1.095\,{\rm GeV}\,,
\\
~ \tilde m_b=m_b\left(\bar\mu\right)=4.47521 \,{\rm GeV}\,.
\label{tom_tilde_mass}
\end{gather}
Finally, the following values are used for the gluon and quark
condensates~\cite{Launer:1983ib,Narison2010,ChenKleivSteeleEtAl2013}:
\begin{gather}
  \glueFourD=(0.075\pm0.02)\ \text{GeV}^4\label{glueFourDValue}, \\
  \glueSixD=((8.2\pm1.0)\ \text{GeV}^2)\label{glueSixDValue} \glueFourD, \\
 \big\langle \overline{q}q \big\rangle = -(0.23 \pm 0.03)^3\ \text{GeV}^3\label{quarkThreeDValue}.
\end{gather}

\section{QCD Laplace Sum-Rules, Analysis, and Results}\label{III}
The QCD correlation function 
$\Pi(q^2)$ satisfies the following dispersion relation
\begin{equation}
\label{DisRel}
    \Pi(q^2) = q^4 \int_{t_0}^\infty \frac{
\rho^{\rm had}(t)
    }{t^2(t-q^2)}dt + a_0+a_1q^2,
\end{equation}
where  
$\rho^{\rm had}(t)$ is the hadronic spectral function, 
the terms with $\{a_0,a_1\}$ represent dispersion-relation subtractions,  and it is understood that $q^2$ is not in the  physical region  $q^2>0$.
\\
The spectral function is given by:
\begin{equation}
\label{SpecFun}   
\rho^{\rm had}(t)    = 2h^2\delta(t-M^2)+\theta\left(t-s_0\right)\frac{1}{\pi}\Im[\Pi(t)],
\end{equation}
where  $M$ is our constituent diquark mass, $h$ is the diquark coupling to 
the current \eqref{Current}, and $s_0$ represents the onset of the QCD continuum contribution $\Im[\Pi(t)]$.

Laplace sum-rules (LSRs) are obtained from \eqref{DisRel} via the Borel transform operator  \cite{Shifman:1978by,Shifman:1978bx}
\begin{equation}
\hat B\equiv 
\lim_{\stackrel{N,~Q^2\rightarrow \infty}{N/Q^2\equiv \tau}}
\frac{\left(-Q^2\right)^N}{\Gamma(N)}\left(\frac{\mathrm{d}}{\mathrm{d}Q^2}\right)^N\quad,
~Q^2=-q^2~.
\label{BorelOp}
\end{equation}
This operator  has the following properties useful to construction of Laplace sum-rules:
\begin{gather}
\hat B\left[a_0+a_1Q^2+\ldots a_m Q^{2m}\right]=0\quad,\quad (m~{\rm finite}),
\label{BorelPoly}\\
 \hat B \left[ \frac{Q^{2n}}{t+Q^2}\right]=\tau \left(-1\right)^nt^n\mathrm{e}^{-t\tau}  \quad,\quad n=0,~1,~2,\ldots ~ 
(n~{\rm finite})\quad .
\label{BorelExp}
\end{gather}
In general, $\hat B$  is related to the inverse Laplace transform  \cite{Bertlmann:1984ih}
\begin{gather}
f\left(Q^2\right)=\int\limits_0^\infty F(\tau) \mathrm{e}^{-Q^2\tau}\, \mathrm{d}\tau\equiv{\cal L}\left[ F(\tau)\right]
~ \Longrightarrow~\frac{1}{\tau}\hat B\left[ f\left(Q^2\right)\right]
=F(\tau)={\cal L}^{-1}
\left[ f\left(Q^2\right)\right]
\label{BorelLaplace}, \\
{\cal L}^{-1}
\left[ f\left(Q^2\right)\right]=\frac{1}{2\pi i}\int\limits_{b-i\infty}^{b+i\infty}
f\left(Q^2\right) \mathrm{e}^{Q^2\tau}\,\mathrm{d}Q^2 \,,
\label{InvLapDef}
\end{gather}
where the real parameter $b$ in the definition \eqref{InvLapDef} of the inverse Laplace transform must be chosen so that $f\left(Q^2\right)$ is analytic to the right of the contour of integration in the complex $Q^2$ plane.

\begin{figure}[hbt]
\centering
\includegraphics[scale=0.17]{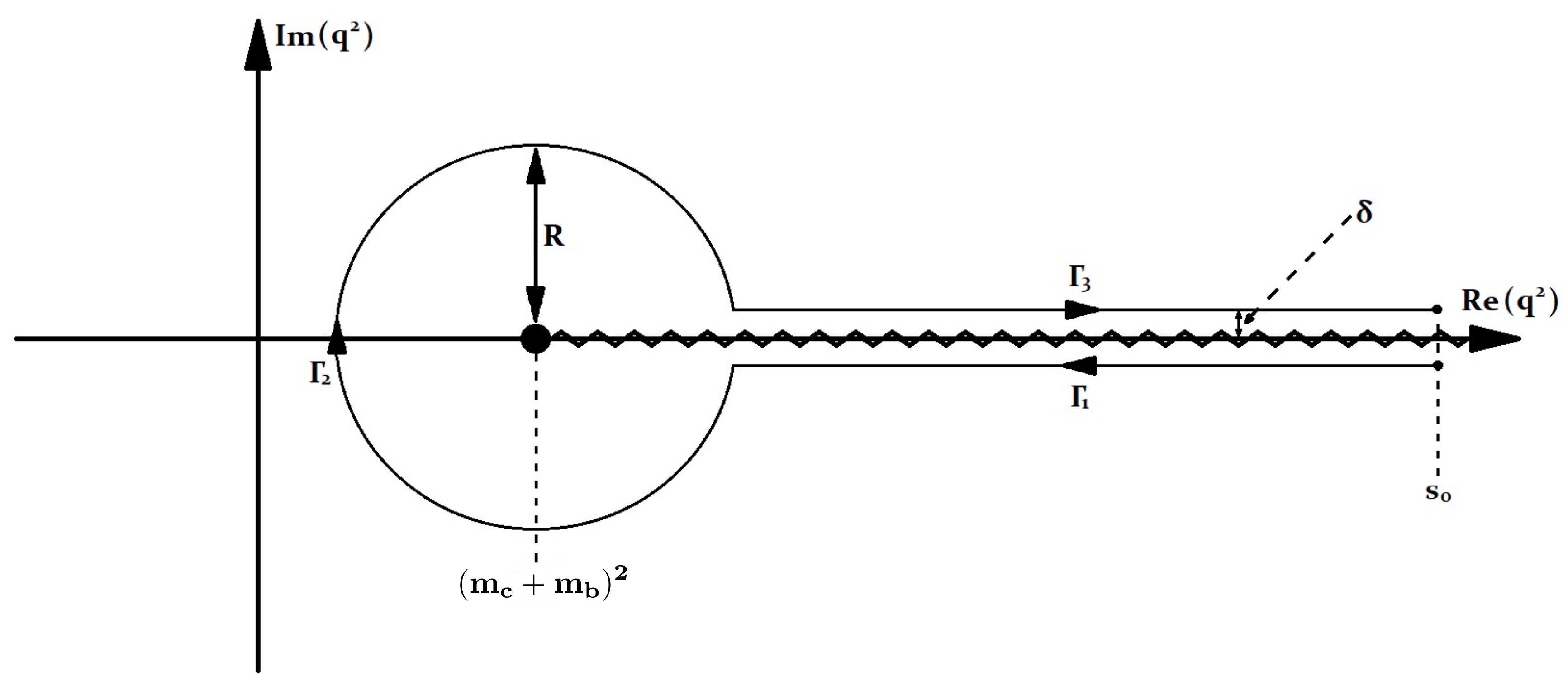}
\caption{The contour of integration used in the evaluation of the LSRs (\ref{LSRControur}).
The contour $\Gamma$ is defined by $\Gamma=\Gamma_1+\Gamma_2+\Gamma_3$.
}
\label{Contour}
\end{figure}

Families of Laplace sum-rules (LSRs) are now obtained by applying $\hat B$ to \eqref{DisRel} weighted by integer powers of $Q^2$, which will involve the theoretically-determined quantity
\begin{equation}
L_k(\tau)\equiv\frac{1}{\tau}\hat B\left[\left(-1\right)^k Q^{2k}\Pi\left(Q^2\right)\right]
={\cal L}^{-1}\left[
\left(-1\right)^k Q^{2k}\Pi\left(Q^2\right)
\right]\,.
\label{laplace}
\end{equation}
For $k\ge 0$, this results in the following Laplace sum-rules relating QCD to hadronic physics
\begin{equation}
L_{k}(\tau)=\int\limits_{t_0}^\infty
t^k \mathrm{e}^{-t\tau}\rho^{\rm had}(t)\,\mathrm{d}t\,,\quad k\ge 0\,.
\label{GenLap}
\end{equation}
Because the $[bc]$ diquark correlator is analytic except for a branch structure  $q^2>\left(m_b+m_c\right)^2$, we can choose $b=0$ for the inverse Laplace transform to find
\begin{equation}
      L_k(\tau)= \frac{1}{2\pi i}\int_{\Gamma}(q^2)^ke^{-q^2\tau}\Pi(q^2) dq^2 +\int_{s_0}^\infty t^k e^{-t\tau} \frac{1}{\pi}\Im{\Pi{(t)}dt}
      +
\frac{1}{2\pi i}\int_{C_\infty}(q^2)^ke^{-q^2\tau}\Pi(q^2) dq^2,
      \end{equation}
where $\Gamma$ is defined in Fig.~\ref{Contour} and $c_\infty$ is the 
$r\to\infty$ limit
of the semi-circular contour 
$q^2=r\exp{\left(i\theta\right)}$ with $\cos\theta >0$. The contribution from $C_\infty$ is zero, and hence 
\begin{equation}
      L_k(\tau)=\frac{1}{2\pi i}\int_{\Gamma}(q^2)^ke^{-q^2\tau}\Pi(q^2) dq^2 +\int_{s_0}^\infty t^k e^{-t\tau} \frac{1}{\pi}\Im{\Pi(t)dt},
 \end{equation}
\\
 which can then be paired with (\ref{SpecFun}) to yield:
  \begin{equation}
 L_k(\tau) = 2h^2M^{2k}e^{-M^2 \tau} + \int_{s_0}^\infty t^k e^{-t\tau} \frac{1}{\pi}\Im{\Pi(t)dt},
 \label{tom_L_k}
 \end{equation}
 where $L_k(\tau)$
 are our unsubtracted  LSRs of integer order $k$ evaluated at Borel scale $\tau$ and $\Gamma$ is the contour as shown in Figure \ref{Contour}. 
By subtracting the QCD continuum contribution 
 \begin{equation}
     \int_{s_0}^\infty t^k e^{-t\tau} \frac{1}{\pi}\Im{\Pi(t)}dt,
 \end{equation}
 from both sides of \eqref{tom_L_k},
 we have the continuum-subtracted LSRs, 
 $\mathcal{R}_k(\tau, s_0)$
\begin{align}
    &\mathcal{R}_k(\tau, s_0) \equiv 
    L_k(\tau)-\int_{s_0}^\infty t^k e^{-t\tau} \frac{1}{\pi}\Im{\Pi(t)}dt=
    \frac{1}{2\pi i}\int_{\Gamma}(q^2)^ke^{-q^2\tau}\Pi^{\text{OPE}}(q^2) dq^2\label{SubLSR}, \\
    \implies& \mathcal{R}_k(\tau, s_0) = 2h^2M^{2k}e^{-M^2 \tau} \,.
    \label{TheoSubLSR}
\end{align}
 The explicit parametrization of each $\Gamma_i$ of $\Gamma$ in (\ref{SubLSR}) is given by
 \begin{equation}
 \label{LSRControur}
     \begin{split}
        \mathcal{R}_k(\tau, s_0) \equiv \frac{1}{2\pi i}\Bigg[&\int_{s_0}^{(m_c+m_b)^2+\sqrt{R^2-\delta^2}}(t-\delta i)^ke^{-(t-\delta i)\tau}\Pi^{\text{OPE}}(t-\delta i) dt  \\ 
     &+\int_{2\pi - \sin^{-1}(\delta/R)}^{\sin^{-1}(\delta/R)}((m_c+m_b)^2+Re^{\theta i} )^ke^{-((m_c+m_b)^2+Re^{\theta i} )\tau}Rie^{\theta i}\Pi^{\text{OPE}}((m_c+m_b)^2+Re^{\theta i} ) d\theta \\
     &+\int_{(m_c+m_b)^2+\sqrt{R^2-\delta^2}}^{s_0}(t+\delta i)^ke^{-(t+\delta i)\tau}\Pi^{\text{OPE}}(t+\delta i) dt
     \Bigg].  
     \end{split}
 \end{equation}
     The above expression was used to numerically calculated the values for $\mathcal{R}_k(\tau, s_0)$. Here, $R$ was set to $(m_c+m_b)^2/2$, and while the 
     ultimate intention is for $\delta \rightarrow 0^+$, $\delta = 10^{-12}\; \text{GeV}^2$ was chosen for numerical calculations.

As described in Section~\ref{II}, pySecDec  \cite{Borowka:2017idc,Borowka:2018goh,Heinrich:2021dbf,Heinrich:2023til} was used to evaluate the master integrals in Eqs.~\eqref{eq:Jints}--\eqref{eq:Bint} which occur in $\Pi\left(q^2\right)$. These integrals are evaluated at discrete $q^2$ points along the contour $\Gamma$
in the upper half plane and then interpolated as needed for numerical integration.  Because pySecDec does not permit direct numerical evaluation for $q^2$ in the lower-half plane,  appropriate symmetry properties   
are applied so that $\Pi\left(q^2\right)$ satisfies Schwarz reflection $\Pi(z^*)=\left[\Pi(z)\right]^*$. 
We find a numerical advantage to choosing the reasonably large value of $R=\left(m_c+m_b\right)^2/2$ because the master integrals vary more slowly along $\Gamma$, decreasing the numerical uncertainty from interpolation.  Finally we note that the numerical determination of 
$\mathcal{R}_k\left(\tau, s_0\right)$ via pySecDec requires numerical inputs of the $\tilde\mu$ in Eq.~\eqref{tom_renorm_scale}  and $\{\tilde m_c,\tilde m_b\}$ in Eq.~\eqref{tom_tilde_mass}. A crucial verification of our determination of 
$\mathcal{R}_k\left(\tau, s_0\right)$ is the 
cancellation of $1/\epsilon$ and $1/\epsilon^2$ divergences within numerical precision.   

 The Laplace sum-rule determination of the  constituent diquark mass $M$ is obtained by taking ratios of (\ref{TheoSubLSR})
     \begin{equation}
         \label{LSRMass}\sqrt{\frac{\mathcal{R}_1(\tau, s_0)}{\mathcal{R}_0(\tau, s_0)}} = M\,.
     \end{equation}
In order to produce values for the mass, a suitable Borel window ($\tau_{\text{min}}, \tau_{\text{max}}$) and $s_0$ value must be determined for each channel.

The process for determining the Borel window begins with the requirement that $\tau_{\text{max}}$ satisfies OPE convergence of the LSRs as $k=0$ as $s_0 \rightarrow \infty$. This is achieved by ensuring that the perturbative contributions, (pert), and the 4d contributions, (4d), of the LSRs obey the inequality 
\begin{equation}
    |\text{pert}| \ge 3\times|4d|
\end{equation}
Furthermore, $\tau_{\text{max}}$ must also satisfy the H\"{o}lder
 inequality condition 
 \begin{equation}
\frac{\mathcal{R}_2(\tau, s_0)/\mathcal{R}_1(\tau, s_0)}{\mathcal{R}_1(\tau, s_0)/\mathcal{R}_0(\tau, s_0)}\ge 1
 \end{equation}
 The above inequality was designed to satisfy the H\"{o}lder inequalities as described in Refs.~\cite{Benmerrouche:1995qa,Kleiv:2013dta}
 for $\mathcal{R}_0(\tau, s_0)$ and $\mathcal{R}_1(\tau, s_0)$.
 
 The lower boundary of the Borel window, $\tau_{\text{min}}$, is constrained by the H\"{o}lder inequality, as discussed above, and by requiring that the resonance contribution ({\it i.e.}, the “pole criterion”) to $\mathcal{R}_1/\mathcal{R}_0$ be 50\% or greater,
 \begin{equation}
\frac{\mathcal{R}_1(\tau, s_0)/\mathcal{R}_0(\tau, s_0)}{\mathcal{R}_1(\tau, \infty)/\mathcal{R}_0(\tau, \infty)}\ge 0.5 \,.
 \end{equation}
The Borel window was determined by maximizing its width while satisfying all of the aforementioned conditions imposed on $\tau_{\text{min}}$ and $\tau_{\text{max}}$.

The value for $s_0$ is selected by minimizing 
\begin{equation}
\label{chiSq}
    \chi^2(s_0, M_{\text{avg}}) = \sum_{j=0}^{20}\left(\frac{1}{M_{\text{avg}}}\sqrt{\frac{\mathcal{R}_1(\tau_j, s_0)}{\mathcal{R}_0(\tau_j, s_0)}} -1 \right)^2 \,,
\end{equation}
where $\left\{\tau_j \right\}_{j=0}^{20}$ is the Borel window divided equally into 20 subintervals, and $M_{\text{avg}}$ is the average mass value given by (\ref{LSRMass}) over the $\tau_j$ values.

The selection of the Borel window and the $s_0$ values are intrinsically linked, such that a change in one affects the other. Both sets of parameters were therefore considered simultaneously, with multiple iterations of the selection algorithm performed until the continuum threshold $s_0$ and the Borel window were optimized and the value of $\chi^2$ in Eq. (\ref{chiSq}) was sufficiently small. We imposed the requirement $\chi^2 < 0.05$.
After the continuum $s_0$ is optimized, the $\tau$-stability criterion
\begin{equation}
\label{massX_v2}
    \frac{d}{d\tau}\sqrt{\frac{\mathcal{R}_1(\tau_j, s_0)}{\mathcal{R}_0(\tau_j, s_0)}} =0 
\end{equation}
is used to extract the final $[bc]$ diquark  mass prediction, consistent  with stability of  the left-hand side of (\ref{LSRMass}) with respect to $\tau$. 

A Monte Carlo uncertainty analysis was performed in which all input parameters were varied within their respective uncertainties over 1000 iterations. This resulted in uncertainties in the predicted masses of less than 10\%. Nevertheless, we conservatively assign a methodological uncertainty of 10\% to the diquark constituent mass values. The mass predictions were found to be most sensitive to variations in the renormalization scale parameter $\bar\mu$ within the region $\overline m_c<\bar\mu<\overline m_b$ . Uncertainties associated with the continuum threshold parameter were estimated by taking the largest deviation from its optimized value that could be accommodated while maintaining the required $\chi^2 < 0.05$ criterion.

\begin{figure}[hbt]
    \centering
    \begin{minipage}{0.45\linewidth}
        \centering
        \includegraphics[width=\linewidth]{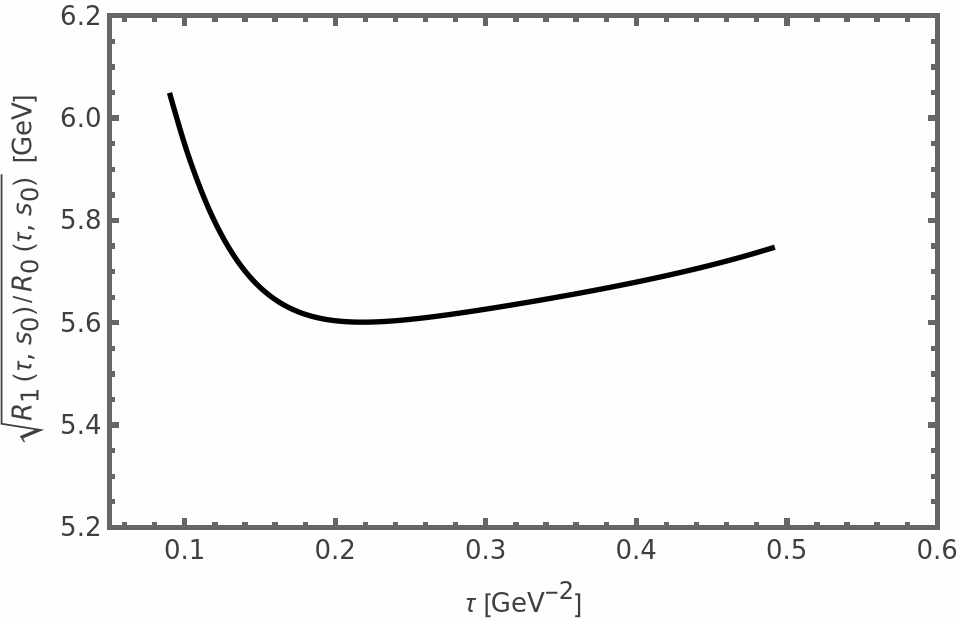}
        \caption{Sum-rule mass-ratio at the optimized continuum threshold parameter $s_0=64\,{\rm GeV^2}$ versus the Borel scale $\tau$ in the axial vector $1^+$ channel.}
        \label{avp}
    \end{minipage}
    \hfill
    \begin{minipage}{0.45\linewidth}
        \centering
        \includegraphics[width=\linewidth]{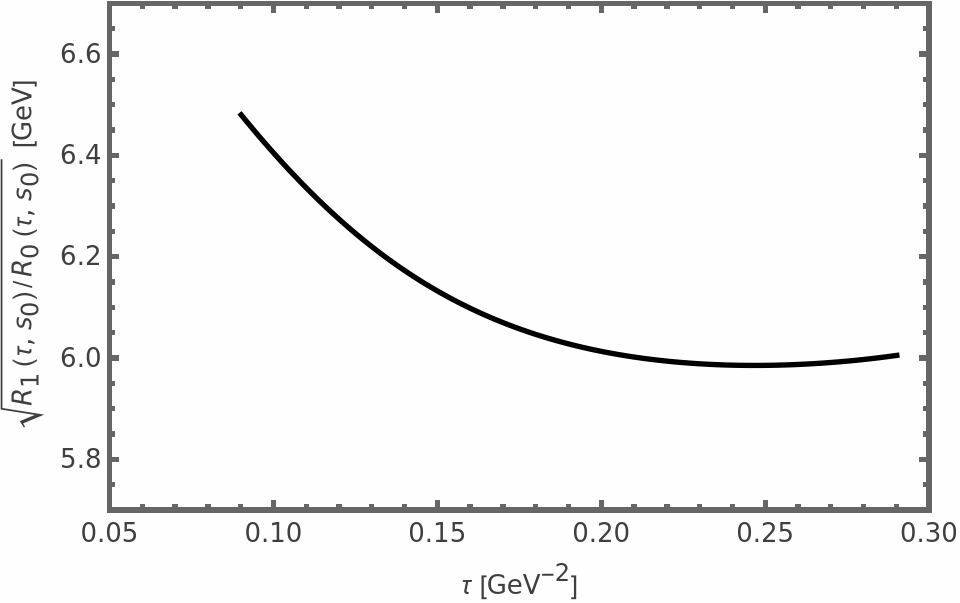}
        \caption{Sum-rule mass-ratio at the optimized continuum threshold parameter $s_0=64\,{\rm GeV^2}$ versus the Borel scale $\tau$ in the vector $1^-$ channel.}
        \label{vp}
    \end{minipage}
    \vspace{1cm}
    
    \begin{minipage}{0.45\linewidth}
        \centering
        \includegraphics[width=\linewidth]{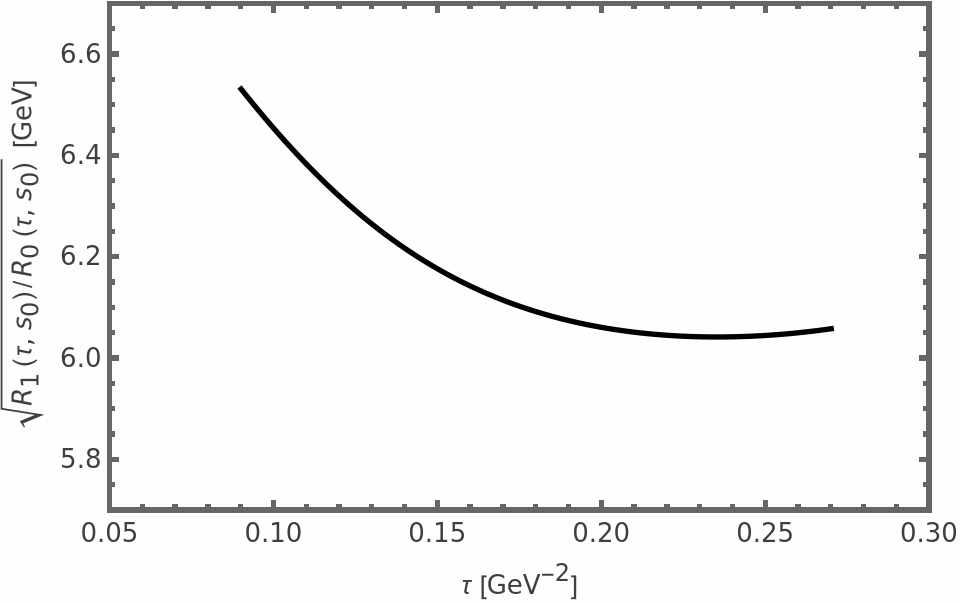}
        \caption{Sum-rule mass-ratio at the optimized continuum threshold parameter $s_0=65\,{\rm GeV^2}$ versus the Borel scale $\tau$ in the pseudoscalar $0^-$ channel.}
        \label{psp}
    \end{minipage}
    \hfill
    \begin{minipage}{0.45\linewidth}
        \centering
        \includegraphics[width=\linewidth]{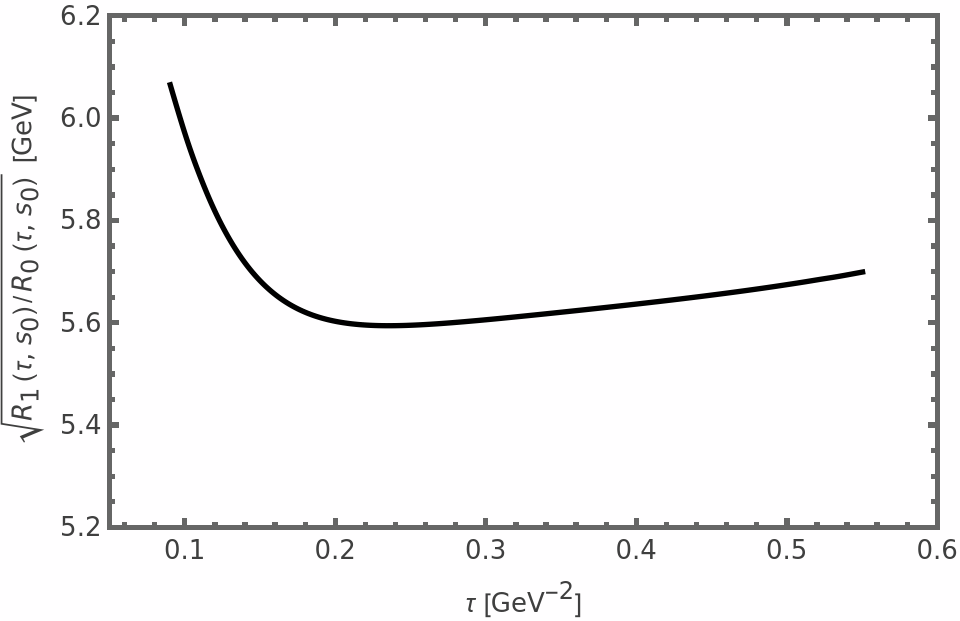}
        \caption{Sum-rule mass-ratio at the optimized continuum threshold parameter $s_0=62\,{\rm GeV^2}$ versus the Borel scale $\tau$ in the scalar $0^+$ channel.}
        \label{sp}
    \end{minipage}
    \end{figure}

Figures (\ref{avp}) through (\ref{sp}) show the sum-rule mass ratio [the left-hand side of Eq.~\eqref{LSRMass}] for the optimized $s_0$ as a function of the Borel parameter $\tau$.  For the central values of the QCD input parameters,  the resulting mass prediction obtained  via the stability criterion  \eqref{massX_v2} is given in Table \ref{MassTable} along with the optimized $s_0$, the associated Borel window, and our estimate of theoretical uncertainties.

\renewcommand{\arraystretch}{1.25}
\begin{table}[hbt]
    \centering
    \begin{tabular}{||c | c | c | c | c||}
    \hline \hline
         $J^P$ & $M$ (GeV) & $s_0$ (GeV$^2$) & $\tau_{\text{min}}$ (GeV$^{-2}$) & $\tau_{\text{max}}$ (GeV$^{-2}$)\\
         
        \hline \hline $0^-$ & 6.04 $\pm$ 0.60 & 65 $\pm$ 11 & 0.04 & 0.27\\
        $0^+$ & 5.60 $\pm$ 0.56 & 62 $\pm$ 10 & 0.04 & 0.55\\
        $1^-$ & 5.99 $\pm$ 0.60 & 64 $\pm$ 10 & 0.04 & 0.29\\
        $1^+$ & 5.60 $\pm$ 0.56 & 64 $\pm$ 12 & 0.04 & 0.49\\
        \hline \hline
    \end{tabular}
    \caption{
  QCD sum-rule predictions and associated uncertainties for the $[bc]$ diquark  constituent mass, optimized continuum threshold  $s_0$, and Borel window are shown for each $J^P$.}
    \label{MassTable}
\end{table}

\section{Discussion  }
\label{IV}

Our QCD sum-rule analysis of the $[bc]$ diquark masses for various $J^P$ values has been performed to NLO in perturbation theory within the $\overline{\rm MS}$ scheme and up to dimension six in QCD condensates.  Our analysis therefore represents an extension of  Ref.~\cite{Tang:2012np}
that worked to LO in perturbation theory, truncated the OPE at dimension four,  and considered only the $0^+$ and $1^+$ quantum numbers.   Taking into account theoretical uncertainties, our mass predictions of Table \ref{MassTable}  are slightly lower than  Ref.~\cite{Tang:2012np}.  

The NLO perturbative calculation is challenging because of the presence of multiple scales (i.e., $m_c$, $m_b$, $q^2$) resulting in highly non-trivial master integrals Eqs.~\eqref{eq:Jints}--\eqref{eq:Bint}. Unfortunately, the analytic expressions for the $\epsilon$-expansions of these integrals are not known to sufficient order, as described earlier. An important feature of our methodology is the use of numerical loop-integration techniques using pySecDec \cite{Borowka:2017idc,Borowka:2018goh,Heinrich:2021dbf,Heinrich:2023til} 
to evaluate these  master integrals on the integration contour of Fig.~\ref{Contour}. Because the radius $R$ is well-separated from the $\left(m_c+m_b\right)^2$ threshold, we find that the master integrals have surprisingly smooth behaviour on the contour compared to the rapid cusp-like behaviour near the threshold.  This provides some numerical advantage when interpolating the master integrals from the discrete pySecDec results. Our pySecDec calculations have been  validated by comparing  limiting cases and  kinematic parameter space regions of the master integrals \eqref{eq:Jints}--\eqref{eq:Bint} where analytic results exist. 

The $[bc]$ diquark constituent masses in Table~\ref{MassTable}  show some interesting physical patterns.  First there is a clear splitting between the positive parity states and  negative parity states as expected from the necessity of the orbital  excitation  needed to achieve negative parity.  Furthermore, there is a near-degeneracy within each parity as expected by the approximate $1/(m_cm_b)$ suppression of the spin-spin interaction in doubly-heavy diquark systems (see e.g., Ref.~\cite{Maiani:2004vq}). Because of this spin-spin suppression caused by the large masses of both quarks (as opposed to diquarks where one quark mass is small),  the $[bc]$ diquarks have the richest structure of quantum numbers suggesting the possibility of a similarly rich structure of tetraquark systems with $[bc]$ diquark content.

Our NLO QCD  Laplace sum-rule analysis of the $[bc]$ diquark constituent mass completes a comprehensive survey of the entire spin-zero and spin-one diquark flavor structure to NLO.  These NLO results are summarized in Tables~\ref{LightDiquarkMassesTable}--\ref{BottomDiquarkMassesTable} along with LO sum-rule analyses. Each of these sum-rule mass predictions can be used as a QCD criterion to assess the constituent diquark mass parameters used in the various diquark substructure models of tetraquark and pentaquark systems  (e.g.,  Type I and Type II diquark models \cite{Maiani:2004vq,Maiani:2005pe,Maiani:2013nmn,Lebed:2016yvr,Maiani:2021tri,Maiani:2014aja}, dynamical quark model \cite{Giron:2021sla,Mutuk:2026bhq},  relativized diquark model \cite{Anwar:2018sol,Ferretti:2020ewe,Bedolla:2019zwg,Anwar:2017toa,Ferretti:2019zyh}, relativistic quark model \cite{Ebert:2005nc,Ebert:2007rn,Ebert:2008kb,Ebert:2010af,Faustov:2021hjs}, Regge relations and string models \cite{Song:2023izj}, diquark effective Hamiltonian model \cite{Shi:2021jyr}).

Our $[bc]$ diquark constituent mass predictions in this paper are  relevant to the $T_{bc}(bc\bar u\bar d)$ tetraquark
mass $m\left(T_{bc}\right)$ and the  question of 
the mass difference from the $\bar B D$ threshold (see e.g., Ref.~\cite{Feng:2026rle} for a literature survey of  predictions indicative of a near-threshold mass
$-50 \,{\rm MeV} <m\left(T_{bc}\right)-m\left(\bar B\right) -m(D) <120\,{\rm MeV}$). The $[bc]$ and $[ud]$ diquark constituent mass parameters have an important role in various models used to study the $T_{bc}$ system, so it is interesting to compare our Table~\ref{MassTable} $[bc]$ mass results with determinations in other models.   The $\{0^+,1^+\}$ diquark masses determined in the relativistic quark model \cite{Ebert:2007rn}, relativized diquark model~\cite{Ferretti:2019zyh},  potential models \cite{Kiselev:2002iy}, and Bethe-Salpeter method \cite{Yu:2018com} are slightly larger than the upper range of  our QCD sum-rule determinations of Table~\ref{MassTable}.  The Regge trajectory analysis of \cite{Feng:2023txx} considers the same  $J^P=\{0^\pm,1^\pm\}$ as our analysis, and although the Regge trajectory masses are again slightly larger, the pattern of mass $\{0^+,1^+\}$ and $\{0^-,1^-\}$ degeneracies 
and the $\{0^+,0^-\}$ $\{1^+,1^-\}$ mass splittings 
are  quite similar to our Table~\ref{MassTable} results.  
Finally, we find excellent agreement of Table~\ref{MassTable} with the $\{0^+,1^+\}$ Regge relation and string model results of Ref.~\cite{Song:2022csw} that are subsequently used in a $T_{bc}$ mass determination in Ref.~\cite{Song:2023izj}. Thus our QCD sum-rule analysis of the $J^P=\{0^\pm,1^\pm\}$ $[bc]$ diquark constituent  masses provides interesting insights on the diquark mass  parameters used in various models of  tetraquark systems.
Furthermore, the  numerical techniques developed in this paper for loop integrations and sum-rule construction provide new methodologies for NLO QCD sum-rule calculations involving multiple scales.  

\section*{Acknowledgements}
T.~de Oliveira is supported by the National Science and Technology Council of Taiwan. S.~Li is supported by the Deutsche Forschungsgemeinschaft (DFG) 
through the Collaborative Research Center 1660 ``Hadrons and Nuclei as Discovery Tools'' (grant 514321794). T.G.~Steele is supported by research funding from the Natural Sciences \& Engineering Research Council of Canada (NSERC Grant SAPIN-2021-00024).


\renewcommand{\arraystretch}{1.25}
\begin{table}[hbt]
    \centering
    \begin{tabular}{|| c || c | c || c | c || c ||}
    \hline \hline 
    $[qq^\prime]$ & \multicolumn{2}{c||}{$[nn]$} & \multicolumn{2}{c||}{$[ns]$} & $[ss]$ \\
    \hline \hline
    $J^P$                     & $0^+$ & $1^+$ & $0^+$ & $1^+$ & $1^+$ \\
    \hline
    $M$ (GeV)  & $0.55$ &        & 0.46     &        & N.S.  \\
    \hline
    $\delta M$ (GeV)  & $0.15$ &  &  &  &  \\
    \hline
    Ref. & \cite{Dosch1989} & & \cite{Zhang:2006xp} & & \cite{deOliveira:2023hma} \\
    \hline 
    Order & NLO & & NLO & & NLO \\
    \hline \hline
    $M$ (GeV)         & 0.50 & 0.90 & 0.70 & 1.00 &  \\
    \hline
    $\delta M$ (GeV)  & 0.20 & 0.20 &  0.20 & 0.20 &  \\
    \hline
    Ref. & \multicolumn{4}{c||}{\cite{Jamin:1989hh}} & \\
    \hline 
    Order & \multicolumn{4}{c||}{NLO} & \\
    \hline \hline
    $M$ (GeV)         & 0.50 & 0.70 & 0.64 & 0.80 &  \\
    \hline
    $\delta M$ (GeV)  & 0.05 & 0.04 &  0.03 & 0.03 &  \\
    \hline
    Ref. & \multicolumn{4}{c||}{\cite{Wang:2011ab}} & \\
    \hline 
    Order & \multicolumn{4}{c||}{NLO} & \\
    \hline \hline
    $M$ (GeV)         & 0.40 & 0.34 & 0.55 & 0.42 &  \\
    \hline
    $\delta M$ (GeV)  &  & 0.04 &  0.03 & 0.03 &  \\
    \hline
    Ref. & \cite{Zhang:2006xp}& \multicolumn{3}{c||}{\cite{Tang:2012np}} & \\
    \hline 
    Order & NLO & \multicolumn{3}{c||}{NLO} & \\
    \hline \hline
       \end{tabular} 
    \caption{Masses of diquarks containing no heavy ($c$ or $b$) quarks determined using QCD sum rules. Here $n\in\{u\,,d\}$, $M\pm\delta M$ is the predicted mass and its uncertainty, and N.S. signifies that the sum rules were not stable, preventing a reliable mass prediction. The ``Order" heading indicates the perturbative order included in the work. Refs.~\cite{Dosch1989,Jamin:1989hh} give ranges for diquark masses; the values quoted above are the central value with half the range as the uncertainty. Ref.~\cite{Zhang:2006xp} does not include uncertainties. Ref.~\cite{Tang:2012np} calculates diquark masses using the sum rules given in Refs.~\cite{Dosch1989,Jamin:1989hh} with different QCD input parameters.}
    \label{LightDiquarkMassesTable}
\end{table}

\renewcommand{\arraystretch}{1.25}
\begin{table}[hbt]
    \centering
    \begin{tabular}{|| c || c | c || c | c || c ||}
    \hline \hline 
    $[cq]$ & \multicolumn{2}{c||}{$[cn]$} & \multicolumn{2}{c||}{$[cs]$} & $[cc]$ \\
    \hline \hline
    $J^P$ & $0^+$ & $1^+$ & $0^+$ & $1^+$ & $1^+$ \\
    \hline
    $M$ (GeV)  & $1.78$ & $1.87$ & $1.82$ & $1.91$ & $3.51$ \\
    \hline
    $\delta M$ (GeV)  & $0.05$ & $0.05$ & $0.08$ & $0.08$ & $0.35$ \\
    \hline
    Ref. & \multicolumn{2}{c||}{\cite{deOliveira:2023hma}} & \multicolumn{2}{c||}{\cite{deOliveira:2023hma}} & \cite{Esau:2019hqw} \\
    \hline 
    Order & \multicolumn{2}{c||}{NLO} & \multicolumn{2}{c||}{NLO} & NLO \\
    \hline \hline
    $M$ (GeV)  & $1.77$ & $1.76$ & $1.84$ & $1.84$ &  $2.99$\\
    \hline
    $\delta M$ (GeV)  & $0.08$ & $0.08$ & $0.08$ & $0.08$ & $0.10$  \\
    \hline
    Ref. & \multicolumn{2}{c||}{\cite{Wang:2010sh}} & \multicolumn{2}{c||}{\cite{Wang:2010sh}} &  \cite{Tang:2012np} \\
    \hline 
    Order & \multicolumn{2}{c||}{LO} & \multicolumn{2}{c||}{LO} &  LO \\
    \hline
    \end{tabular} 
    \caption{Masses of diquarks containing at least one $c$ quark determined using QCD sum rules. All notations are identical to those in Table~\ref{LightDiquarkMassesTable}. Note that the $J^P=0^+,1^+$ masses were also determined in Ref.~\cite{Kleiv:2013dta}. Subsequently the analysis was extended and refined in Ref.~\cite{deOliveira:2023hma}, so we only quote its results here. The Heavy-strange mass given for Ref.~\cite{deOliveira:2023hma} is for $\kappa=\frac{\langle\bar ss\rangle}{\langle\bar nn\rangle}=0.74$.}
    \label{CharmDiquarkMassesTable}
\end{table}

\begin{table}[hbt]
    \centering
    \begin{tabular}{||c || c | c || c | c || c | c | c | c || c ||}
    \hline \hline 
    $[bq]$ & \multicolumn{2}{c||}{$[bn]$} & \multicolumn{2}{c||}{$[bs]$} & \multicolumn{4}{c||}{$[bc]$} & $[bb]$ \\
    \hline \hline
    $J^P$ & $0^+$ & $1^+$ & $0^+$ & $1^+$ & $0^+$ & $1^+$ & $0^-$ & $1^-$ & $1^+$ \\
    \hline
    $M$ (GeV)  & $4.97$ & $4.97$ & $5.04$ & $5.05$ & $5.60$ & $5.60$ & $6.04$ & $5.99$ & $8.67$ \\
    \hline
    $\delta M$ (GeV)  & $0.08$ & $0.08$ & $0.11$ & $0.11$ & $0.56$ & $0.56$ & $0.60$ & $0.60$ & $0.69$ \\
    \hline
    Ref. & \multicolumn{2}{c||}{\cite{deOliveira:2023hma}} & \multicolumn{2}{c||}{\cite{deOliveira:2023hma}} & \multicolumn{4}{c||}{This Work} & \cite{Esau:2019hqw} \\
    \hline 
    Order & \multicolumn{2}{c||}{NLO} & \multicolumn{2}{c||}{NLO} & \multicolumn{4}{c||}{NLO} & NLO \\
   \hline \hline
    $M$ (GeV)  & $5.14$ & $5.13$ & $5.20$ & $5.20$ & $6.30$ & $6.36$ & &  & $9.76$ \\
    \hline
    $\delta M$ (GeV)  & $0.12$ & $0.11$ & $0.07$ & $0.08$ & $0.09$ & $0.08$ & & & $0.08$  \\
    \hline
    Ref. & \multicolumn{2}{c||}{\cite{Wang:2010sh}} & \multicolumn{2}{c||}{\cite{Wang:2010sh}} & \multicolumn{2}{c|}{\cite{Tang:2012np}} & & & \cite{Tang:2012np} \\
    \hline 
    Order & \multicolumn{2}{c||}{LO} & \multicolumn{2}{c||}{LO} & \multicolumn{2}{c|}{LO} & & & LO \\
   \hline \hline
   \end{tabular} 
    \caption{Masses of diquarks containing at least one $b$ quark determined using QCD sum rules. All notations and parameter values are identical to those in Table~\ref{CharmDiquarkMassesTable}. }
    \label{BottomDiquarkMassesTable}
\end{table}

\clearpage

\bibliographystyle{h-physrev}
\bibliography{main}

\end{document}